\documentclass[11pt,reqno,preprint]{article}
\usepackage{jheppub,epsfig,amssymb,amsmath,mathrsfs,hyperref,array,hhline}
\usepackage[makeroom]{cancel}
\usepackage{multirow,bbm}
\usepackage{feynmp-auto}

\usepackage{todonotes}
\usepackage{mathtools}
\usepackage{caption}
\usepackage{subcaption}
\usepackage{placeins}

\makeatletter
\DeclareRobustCommand*{\bfseries}{%
  \not@math@alphabet\bfseries\mathbf
  \fontseries\bfdefault\selectfont
  \boldmath
}

\def\e{\epsilon}
\def\be{\begin{equation}}
\def\ee{\end{equation}}
\newcommand{\bea}{\begin{eqnarray}}
\newcommand{\eea}{\end{eqnarray}}

\def\Fig#1{Fig.~{\ref{#1}}}

\def\eqn#1{eq.~\eqref{#1}}

\def\blue#1{{\color{blue}#1}}
\def\green#1{{\color{green}#1}}

\def\lr{\leftrightarrow}

\def\Gcusp{\Gamma_{\rm cusp}}

\def\Li{\textrm{Li}}

\def\EE{{\cal E}}

\newcommand{\del}{\partial}

\newcommand{\cC}{\mathcal{C}}

\newcommand{\cM}{\mathcal{M}}

\newcommand{\cO}{\begin{cal}O\end{cal}}

\newcommand{\cS}{\begin{cal}S\end{cal}}

\newcommand{\fwboxL}[2]{\text{\makebox[#1][l]{$#2$}}}

\DeclareMathOperator{\tr}{tr}

\def\blue#1{{\color{blue}#1}}

\title{A Three-Point Form Factor Through Five Loops}

\author{Lance~J.~Dixon$^{1}$, Andrew~J.~McLeod$^{2}$ and Matthias Wilhelm$^{2}$}

\affiliation{$^1$ SLAC National Accelerator Laboratory,
Stanford University, Stanford, CA 94309, USA}

\affiliation{$^2$ Niels Bohr International Academy, Niels Bohr Institute,
  Blegdamsvej 17, 2100 Copenhagen \O{}, Denmark}

\abstract{We bootstrap the three-point form factor
  of the chiral part of the stress-tensor supermultiplet in planar
  $\mathcal{N}=4$ super-Yang-Mills theory, obtaining new results at three,
  four, and five loops.  Our construction employs known conditions on
  the first, second, and final entries of the symbol, combined with
  new multiple-final-entry conditions, ``extended-Steinmann-like'' conditions,
  and near-collinear data from the recently-developed form factor
  operator product expansion. 
  Our results are expected to give the
  maximally transcendental parts 
  of the $gg\to Hg$ and $H\to ggg$
  amplitudes in the heavy-top limit of QCD.
  At two loops, the extended-Steinmann-like 
  space of functions we describe contains all 
  transcendental functions required for four-point 
  amplitudes with one massive and three 
  massless external legs, and all massless 
  internal lines, including processes such as 
  $gg\to Hg$ and $\gamma^*\to q\bar{q}g$. 
  We expect the extended-Steinmann-like 
  space to contain these amplitudes 
  at higher loops as well, 
  although not to arbitrarily high loop order.
  We present evidence that the planar $\mathcal{N}=4$ three-point form factor
  can be placed in an even smaller space of functions, with no independent
  $\zeta$ values at weights two and three.}

\preprint{ \begin{flushright} SLAC--PUB--17581 \end{flushright}}

\begin{document}
\maketitle
\flushbottom
\begin{fmffile}{feyndiags}


\section{Introduction}
\label{sec:introduction}

The most important production mechanism for the Higgs boson at the
Large Hadron Collider is
the gluon fusion process, mediated by a top quark loop.  Higher-order
QCD corrections to this process are very large, necessitating
its understanding to at least next-to-next-to-next-to-leading
order~(N$^3$LO) in the strong coupling
$\alpha_s$~\cite{Anastasiou:2015vya,Mistlberger:2018etf}.
In particular, matrix elements for the Higgs boson plus $n$ additional gluons
are required.  In the limit where the mass of the top quark is taken to be
infinite, these amplitudes can be thought of as the $n$-point form factors
of the gauge-invariant local composite operator
$\tr(F^2)$~\cite{Wilczek:1977zn,Shifman:1978zn}:
\begin{equation}
  \mathcal{F}_{\tr(F^2)}(p_1,\dots,p_n;q)
  = \int d^dx \, e^{-iq\cdot x}\langle1,\dots,n|\tr(F^2)(x)|0\rangle\,,
\end{equation}
where $p_i$ denotes the momentum of the $i^{\text{th}}$ external particle,
and $q$ is the momentum of the operator insertion, or equivalently
the momentum of the Higgs boson.

At tree level, the operator $\tr(F^2)$ can be split into a self-dual part
$\tr(F_{SD}^2)$ and an anti-self-dual part $\tr(F_{ASD}^2)$,
exposing a helicity structure similar to that for pure-gluon
amplitudes~\cite{Dixon:2004za}.%
\footnote{In particular, the form factor of $\tr(F_{ASD}^2)$ can be obtained from the form factor of $\tr(F_{SD}^2)$ by parity, which acts on the external states as well as on the operator.}
At loop level, when the form factor is normalized by its tree-level value,
the kinematic dependence is trivial for $n=2$
(where it corresponds to the Sudakov form factor),
and becomes non-trivial for $n\geq3$.
The three-point form factor $\mathcal{F}_{\tr(F^2)}$ in QCD is currently known
up to two loops~\cite{Gehrmann:2011aa}, which is the order contributing to the
N$^3$LO terms in the Higgs production cross section.  Curiously, the
maximally transcendental part of this form factor, or more specifically
of its remainder function~\cite{Bern:2008ap,Drummond:2008aq},
has been shown to coincide with its counterpart in the maximally
supersymmetric Yang-Mills ($\mathcal{N}=4$ sYM)
theory~\cite{Brandhuber:2012vm}, where $\tr(F_{SD}^2)$ is part of the larger
chiral part of the stress-tensor supermultiplet.
This finding is an example of the more general principle of maximal
transcendentality
\cite{Kotikov:2001sc,Kotikov:2002ab,Kotikov:2004er,Kotikov:2007cy},
which relates the results for many interesting quantities in
$\mathcal{N}=4$ sYM theory to the maximally transcendental part of
their counterparts in pure Yang-Mills theory.%
\footnote{Notably, this principle does not hold for massless
scattering amplitudes, starting with the four-gluon amplitude.}

Form factors of a variety of operators have been
calculated in $\mathcal{N}=4$ sYM theory using modern techniques originally developed
in the context of scattering amplitudes.
These include recursion relations~\cite{Brandhuber:2010ad,Brandhuber:2011tv,Bolshov:2018eos, Bianchi:2018peu}, on-shell diagrams, polytopes and Gra\ss{}mannians~\cite{Bork:2014eqa, Bork:2015fla,Frassek:2015rka, Bork:2016hst, Bork:2016xfn, Bork:2017qyh}, twistor space actions~\cite{Koster:2016ebi, Koster:2016loo, Chicherin:2016qsf, Koster:2016fna}, a connected prescription~\cite{He:2016dol, Brandhuber:2016xue, He:2016jdg}, color-kinematics duality \cite{Boels:2012ew, Yang:2016ear,Lin:2020dyj}, and a dual description via the AdS/CFT correspondence \cite{Alday:2007he,Maldacena:2010kp,
  Brandhuber:2010ad,Gao:2013dza,Ben-Israel:2018ckc,Bianchi:2018rrj}. 
These results are currently limited to two loops for $n\geq3$.  
See ref.\ \cite{Yang:2019vag} for a recent review.

Prominent among these modern techniques are bootstrap
methods~\cite{Dixon:2011pw,Dixon:2013eka,Caron-Huot:2016owq},
which have been used in planar $\mathcal{N}=4$ sYM theory to
determine the six-point amplitude through
seven loops \cite{Caron-Huot:2019vjl} and the
seven-point amplitude through four
loops~\cite{Drummond:2014ffa,Dixon:2016nkn,Drummond:2018caf,Dixon:2020cnr}.
They have also been used to verify~\cite{Almelid:2017qju} the three-loop
soft anomalous dimension matrix~\cite{Almelid:2015jia},
which contributes to the infrared structure of 
gauge theory beyond the planar limit.
Bootstrap methods are based on making an ansatz for the functional form of the
amplitude, leveraging knowledge of the location of its possible
discontinuities, as well as conditions on its derivatives and
higher discontinuities. Coefficients in the ansatz are fixed by requiring
it to have appropriate discrete symmetries
and by matching it to various physical limits.

One important set of constraints for these bootstrap methods comes from the behavior of these amplitudes in 
near-collinear limits. Their expansion around this limit is described at
all loop orders by the integrability-based pentagon operator product 
expansion (POPE)~\cite{%
  Basso:2013vsa, Basso:2013aha,Basso:2014koa,Basso:2014nra,Basso:2014hfa,%
  Basso:2015rta,Basso:2015uxa,Belitsky:2014sla,Belitsky:2014lta,%
  Belitsky:2016vyq}. 
Recently, an analogous description of the near-collinear limit of form factors
has been developed, termed the form factor operator product expansion
(FFOPE)~\cite{Sever:2020jjx,Toappear1,Toappear2}. It provides physical
constraints on the near-collinear behavior of the form factors of the
chiral part of the stress-tensor supermultiplet at any loop order.

In this paper, we study the three-point maximum-helicity-violating (MHV) form factor of 
the chiral part of the stress-tensor supermultiplet. This form factor depends on the
three ratios $u=s_{12}/q^2$, $v=s_{23}/q^2$ and $w=s_{31}/q^2$.  However, since
$q^2 = s_{123} = s_{12} + s_{23} + s_{31}$, these variables satisfy the
constraint $u+v+w=1$, and the form factor actually depends on only two variables.
These are exactly the same variables on which the corresponding 
Higgs boson amplitudes in QCD depend.  The two-loop contribution to this 
form factor has been computed via unitarity methods;
moreover, it was shown to be uniquely determined by a bootstrap
approach involving conditions on the first, second, and final entries
of the symbol~\cite{Brandhuber:2012vm}.  The first two entry conditions
restrict the single and double discontinuities of the form factor,
while the final-entry conditions constrain its differential structure. 
Here we exploit the recent progress on the FFOPE to bootstrap this form factor
all the way to five loops. 

In addition to leveraging the FFOPE, we observe and exploit two 
new forms of mathematical structure in the infrared-finite
part of the form factor, which are not obeyed by the remainder function.
First, we observe new multiple-final-entry conditions that
restrict linear combinations of the last two or three entries
of the symbol of the form factor.
Second, we discover certain ``extended-Steinmann-like'' (ES-like)
conditions~\cite{Caron-Huot:2019bsq,Drummond:2017ssj} at all depths in the symbol. 
In the three-point form factor alphabet $\{ u,v,w,1-u,1-v,1-w\}$, these ES-like conditions
imply that $1-u$ never appears next to $1-v$ or $1-w$ in the symbol, nor $1-v$ next to $1-w$.
While this condition does not transparently follow from the standard Steinmann relations, it
is inspired by studying the Steinmann and cluster algebra constraints for
heptagon functions~\cite{Dixon:2016nkn,Drummond:2017ssj}, since the pentabox ladder
integrals~\cite{Drummond:2010cz,Caron-Huot:2018dsv} contribute to both the seven-particle amplitude
and the three-point form factor.

To carry out our bootstrap, we define a space of polylogarithmic functions $\cM$ that contains the infrared-finite part of the
form factor. This function space draws from the symbol alphabet
$\{ u,v,w,1-u,1-v,1-w\}$ and obeys the ES-like conditions described in the last paragraph.
We find that the dimension of $\cM$ is {\it exactly} $3^w$ at weight $w$ (at symbol level).  Although this growth rate is considerably slower than without the ES-like conditions, it is still considerably faster than the hexagon function space ${\cal H}$ associated with the six-point amplitude, which grows approximately as $\sim 1.8^w$~\cite{Caron-Huot:2019bsq}. On the other hand, the simple dependence of
the dimension on the weight suggests that a direct construction of this
function space should be possible.  

In the absence of a direct construction of $\cM$, it would be helpful if there were a smaller space
of functions that contained the finite part of the form factor. And indeed, when we normalize the form 
factor in a slightly different way, we learn
by taking its derivatives (or rather, its iterated $\{n-1,1\}$ coproducts)
that the space $\cM$ is still larger than necessary for the
planar ${\cal N}=4$ form factor. We also begin to see aspects
of cosmic Galois theory, or the coaction
principle~\cite{Schnetz:2013hqa,Brown:2015fyf,Panzer:2016snt,Schnetz:2017bko},
which is also seen in the
space of hexagon functions~\cite{Caron-Huot:2019bsq}.
Notably, in this smaller space $\cC$,
the constants $\zeta_2$ and $\zeta_3$ no longer
need to be treated as independent functions; they are locked to the other,
symbol-level functions.

The rest of this paper is structured as follows.
We review some basic properties of the three-point MHV form factor of
the chiral part of the stress tensor in section \ref{sec:background}.
Moreover, we analyze its two-loop remainder function, which exhibits many
of the properties that we will generalize to higher loop orders in subsequent
sections.
In section \ref{sec:func_space}, we construct the function space $\cM$
in which the form factor lives. 
We then bootstrap this form factor at three-, four-, and five-loop orders, and analyze the minimal space 
$\cC \subset \cM$ that appears in the coproduct of these functions in section \ref{sec:bootstrapping}.
In section \ref{sec:results}, we study the behavior of the remainder
function through five loops, plotting its dependence on various combinations
of parameters, and considering several kinematic limits.
In section \ref{sec:generalintegrals},
we present evidence that the space $\cM$ may also
govern amplitudes with the same kinematics in arbitrary massless theories.
Our conclusions and outlook are contained in section \ref{sec:conclusions}.

There are two appendices. Appendix~\ref{app:pentaladders}
describes the pentabox ladders and their relation to both $\cM$
and heptagon functions.  Appendix~\ref{app:OPEformulas}
collects some explicit results for the near-collinear limits of the
form factor remainder function needed to make contact with the FFOPE.

{\bf Ancillary files}:  We include three ancillary files.
The first one, {\tt cEandRsymbols.txt},
gives the symbols of the finite part of the form factor through five loops,
and of the remainder function through four loops.
The second and third files, {\tt T2terms.txt} and {\tt T4terms.txt},
provide the $T^2$ and $T^4$ terms, respectively,
in the near-collinear limit through five loops.
%


\section{BPS Form Factors and Polylogarithms}
\label{sec:background}

In this paper, we study the half-BPS operator corresponding to the chiral part 
of the stress-tensor supermultiplet in planar $\mathcal{N}=4$ sYM theory. 
This supermultiplet includes the scalar operator $\tr(\phi^2)$, 
which is part of the so-called $20^\prime$ multiplet,
as well as the chiral part of the on-shell Lagrangian, 
which includes the self-dual operator $\tr(F_{SD}^2)$. 
We refer to refs.~\cite{Eden:2011yp,Brandhuber:2011tv,Bork:2014eqa}
for more background  on these form factors, and their formulation in
$\mathcal{N}=4$ harmonic superspace.

Similar to scattering amplitudes, a helicity degree can be assigned to form 
factors, corresponding to the helicity of the $n$ external massless states. 
Form factors with helicity degree $n-k-2$ are related to those with helicity 
degree $k$ by parity, which acts on the external states as well as
on the operator. Thus, non-trivial N$^k$MHV form factors first occur 
for $n=2k+2$. 

For $n=2$ external states, the form factor only depends on the single scale
$s_{12}=(p_1+p_2)^2$, where $p_i$ is the momentum associated with the
$i^\text{th}$ external state. Dimensional analysis dictates that this
scale dependence has to factor out, leaving us with just a number at
each perturbative order. The first case described by a non-trivial
\emph{function} thus involves $n=3$ external states.
The only non-trivial form factor for this number of external states is MHV,
namely $\mathcal{F}_{n,k=0}\equiv \mathcal{F}_n^{\rm MHV}$.

The form factor $\mathcal{F}_3^{\rm MHV}$ is a function of the three Mandelstam
variables $s_{12}$, $s_{23}$, $s_{31}$ with 
\begin{align}
s_{ij} = (p_i+p_j)^2 = 2p_i\cdot p_j\,.
\label{sijdef}
\end{align}
We rescale these invariants by the invariant mass of the operator's momentum,
\be
q^2 = s_{123} = (p_1+p_2+p_3)^2 = s_{12} + s_{23} + s_{31} \,,
\label{s123def}
\ee
in order to define three dimensionless ratios,
\be
u = \frac{s_{12}}{s_{123}} \,, \quad
v = \frac{s_{23}}{s_{123}} \,, \quad
w = \frac{s_{31}}{s_{123}} \,.
\label{uvwdef}
\ee
These variables obey the constraint
\be
u + v + w = 1 \,
\label{uvwsumto1}
\ee
due to momentum conservation;
as in the case $n=2$, the overall scale dependence factors out.

The form factor $\mathcal{F}_{n}^{\rm MHV}$
obeys an exponentiation relation similar to that of amplitudes in planar $\mathcal{N}=4$ sYM theory~\cite{Brandhuber:2012vm}.
We thus define a finite remainder function by
\begin{equation}
 \mathcal{F}_{n}^{\rm MHV} = \mathcal{F}_n^{\text{BDS}} \exp[ R_n ] \,,
\label{remainderdef}
\end{equation}
where $\mathcal{F}_n^{\text{BDS}}$ is an appropriately
exponentiated one-loop form factor~\cite{Bern:2005iz}. 
(We will provide more details in the three-point case shortly.)
Because the collinear behavior of amplitudes iterates in planar
${\cal N}=4$ sYM theory~\cite{Anastasiou:2003kj},
the remainder function has smooth collinear limits, 
\begin{equation}
R_n \xrightarrow[]{p_i || p_{i+1}} R_{n-1} \,.
\end{equation}
Moreover, it is only non-zero starting at three points and at two loops,
when expanded in the 't Hooft coupling for gauge group $SU(N)$,
$\smash{g^2 = \frac{g^2_{\text{YM}} N}{16 \pi^2}}$:
\begin{align}
R_n = \sum_{L=2}^\infty g^{2L} \, R^{(L)}_n \,.
\end{align}
Thus, the three-point form factor remainder function has vanishing
collinear limits in all three channels,
\begin{equation}
R_3 \xrightarrow[]{p_i || p_{i+1}} 0 \,.
\end{equation}
%

\subsection{The two-loop remainder function}

The two-loop contribution to $\mathcal{F}_{3}^{\rm MHV}$ was computed explicitly
in~ref.~\cite{Brandhuber:2012vm}. There, it was found that the
remainder function could be
expressed entirely in terms of classical polylogarithms as
\begin{align} \label{eq:two_loop_remainder}
  R^{(2)}_3 & =  -2 \sum_{i=1}^3 \left[
    \mathrm{J}_4 \left( -\frac{u_i u_{i+1}}{u_{i+2}}\right)
    + 4 \, \mathrm{Li}_4 \left(1-1/u_i\right)+\frac{\ln^4 u_i}{3!} \right]
  -\frac{\ln^4(u v w)}{4!}  \nonumber\\
&\qquad -2 \left[ \sum_{i=1}^3 \mathrm{Li}_2 (1-1/u_i) \right]^2
  +\frac{1}{2} \left[ \sum_{i=1}^3 \ln^2 u_i\right]^2
  - \frac{23}{2} \zeta_4   \  ,  
\end{align}
where $u_{i+3} = u_i$, $\{u_1, u_2, u_3\} = \{u,v,w\}$, and we have made
use of the function
\be
\mathrm{J}_4(t) = \mathrm{Li}_4(t)-\ln(-t) \mathrm{Li}_3(t)+\frac{\ln^2(-t)}{2!} \mathrm{Li}_2(t)-\frac{\ln^3(-t)}{3!} \mathrm{Li}_1(t) - \frac{\ln^4(-t)}{48} \ .
\ee
While this form of $R^{(2)}_3$ makes its dihedral symmetry manifest, we recall that it is really just a function of two variables due to the constraint~\eqref{uvwsumto1}.

Classical polylogarithms are particular examples of generalized
polylogarithms, or iterated integrals over logarithmic integration
kernels~\cite{%
  Chen,G91b,Goncharov:1998kja,Remiddi:1999ew,Borwein:1999js,Moch:2001zr}.
These functions can be defined iteratively by an integration base point
and their total differential,
\be \label{eq:total_diff}
dF = \sum_\phi F^\phi \, d \ln \phi \, ,
\ee
where the sum is over all logarithmic branch points appearing in the polylogarithmic function $F$. They are commonly expressed in the notation
\be \label{eq:G_func_def}
G_{a_1,\dots, a_n}(z) = \int_0^z \frac{dt}{t-a_1} G_{a_2,\dots, a_n}(z)\,, \qquad G_{\fwboxL{27pt}{{\underbrace{0,\dots,0}_{p}}}}(z) = \frac{\ln^p z}{p!} \,.
\ee
For instance, classical polylogarithms become 
\be
\Li_n(z) = -G_{\fwboxL{34pt}{{\underbrace{0,\dots,0}_{n-1},1}}}(z) \, .
\ee
Another important subclass of generalized polylogarithms are the harmonic polylogarithms (HPLs)~\cite{Remiddi:1999ew} $H_{\vec{a}}(z)$ with $a_i \in \{0,1,-1\}$. They are related by
\begin{equation}
 H_{\vec{a}}(z)=(-1)^p G_{\vec{a}}(z)\,,
\end{equation}
where $p$ is the number of letters $1$ in $\vec{a}$.
Transcendental constants such as multiple zeta values (MZVs)
and alternating Euler-Zagier sums also naturally appear in this space,
as special values of the functions~\eqref{eq:G_func_def}.

Generalized polylogarithms can be assigned a transcendental weight,
corresponding to the number of logarithmic integrations that appear
in their definition; the weight of a product of polylogarithms
is given by the sum of weights. In the $G_{a_1,\dots, a_n}(z)$ notation,
this simply corresponds to the number of indices $n$. Similar to amplitudes
in planar $\mathcal{N}=4$ sYM theory, the form factor remainder function~\eqref{eq:two_loop_remainder} is observed to have a uniform
transcendental weight equal to twice the loop order.

To understand the analytic structure of the polylogarithmic function~\eqref{eq:two_loop_remainder}, it proves useful to study its
symbol~\cite{Goncharov:2010jf}. The symbol of a generic polylogarithm
$F$ can be defined recursively in terms of its total differential~\eqref{eq:total_diff} to be
\be
\mathcal{S} \left(F  \right) = \sum_\phi \mathcal{S}(F^\phi) \otimes \phi \, ,
\label{SymbolDef}
\ee
where $\mathcal{S}$ maps a rational function back to itself, allowing this
recursion to terminate.
The symbol can also be defined as the maximal iteration of the
motivic coaction that acts on generalized polylogarithms~\cite{%
  Gonch3,FBThesis,Brown1102.1312,Duhr:2012fh,2015arXiv151206410B}.
This definition allows one to retain information about all constant letters
$\phi$ other than $i \pi$, which can be useful in practice;
see for instance ref.~\cite{Bourjaily:2019igt}.

The algebraic functions $\phi$ that appear in the tensor
product~(\ref{SymbolDef})
are referred to as symbol letters, and the set of all multiplicatively
independent letters appearing in the symbol of $F$ is referred to as
its symbol alphabet. The symbol alphabet identifies the location of a
function's logarithmic branch points; correspondingly, the symbol can
be thought of as extracting all of a function's non-zero sequences of
discontinuities. 

The symbol of $R_3^{(2)}$ is found to take the remarkably simple
form~\cite{Brandhuber:2012vm}:
\begin{eqnarray}
\cS \Big( R_3^{(2)} \Big) & = &
 4 \, \bigg[ -2 u\otimes (1-u)\otimes (1-u)\otimes \frac{1-u}{u}
+u\otimes (1-u)\otimes u\otimes \frac{1-u}{u}
\nonumber \\
&&\quad 
   -u\otimes (1-u)\otimes v\otimes \frac{1-v}{v}
   -u\otimes (1-u)\otimes w\otimes \frac{1-w}{w}
      \nonumber \\
   &&\quad 
   -u \otimes v \otimes (1-u)\otimes \frac{1-v}{v}
   -u \otimes v \otimes (1-v)\otimes \frac{1-u}{u}
   \nonumber \\
   &&\quad 
   +u\otimes v \otimes w \otimes \frac{1-u}{u}
   +u\otimes v \otimes w \otimes \frac{1-v}{v}
   \nonumber \\
   &&\quad 
      +u\otimes v\otimes w\otimes \frac{1-w}{w}
   -u\otimes w\otimes (1-u)\otimes\frac{1-w}{w}
\nonumber \\
&&\quad 
   +u\otimes w\otimes v\otimes \frac{1-u}{u}
   +u\otimes w\otimes v\otimes \frac{1-v}{v}
\nonumber \\
&&\quad 
   +u\otimes w\otimes v\otimes \frac{1-w}{w}
   -u\otimes w\otimes (1-w)\otimes
   \frac{1-u}{u} \bigg] \ + \ \text{cyclic} \,,
   \label{R32symbol}
   \end{eqnarray}
where the expression in the brackets is summed over all three cyclic
permutations ($u_i \to u_{i+1}$) of the external states.
From this result, we can read off the symbol alphabet of $R_3^{(2)}$ to be
\be
\mathcal{S}_3 = \{ u,\, v,\, w, \, 1-u, \, 1-v, \, 1-w \} \, .
\label{alphabet}
\ee
In addition, we see that the first entry of the symbol is always drawn
from the smaller set
\be
\{ u,\, v,\, w \}\, ,
\label{firstentry}
\ee
consistent with the branch cuts for massless processes always
starting at either $s_{ij}=0$ or $s_{123}=0$. 

There is a natural action of the dihedral group $D_3 \equiv S_3$,
which is generated by the two transformations:
\begin{equation}
 \begin{aligned}
 \hbox{cycle:}\quad  &u \to v \to w \to u,\\
\hbox{flip:}\quad &u \lr v.
\end{aligned}
\label{eq:dihedral}
\end{equation}
The MHV form factor and remainder function should both be invariant
under $S_3$, i.e.~under all permutations of $u,v,w$.

After eliminating $w$ in favor of $u$ and $v$ in the symbol
alphabet~\eqref{alphabet} using the constraint~\eqref{uvwsumto1}, the symbol alphabet can be
rewritten as 
\be
\mathcal{S}_3 = \{ u,\, v, \, 1-u, \, 1-v, \, u+v, \, 1-u-v \} \, .
\label{alphabet2}
\ee
It follows that not all functions that draw on this symbol alphabet can
be written as (products of) single-variable functions. Instead, the
alphabet~\eqref{alphabet2} can be seen to give rise to the
space of 2dHPLs~\cite{Gehrmann:2000zt}. 

\subsection{Adjacent-letter restrictions on the form factor}
\label{adjrestrict}

Further interesting properties arise when we study the finite part of
the form factor itself, instead of the remainder function.
This procedure is similar to defining a BDS-like normalized amplitude,
which exposed the constraints from the Steinmann relations in
six- and seven-point amplitudes~\cite{Caron-Huot:2016owq,Dixon:2016nkn}.
The one-loop form factor is~\cite{Brandhuber:2010ad,Brandhuber:2012vm}
\be
M_3^{(1)}(\e,s_{ij})
= - \frac{1}{\e^2} \sum_{i=1}^3 \left(\frac{\mu^2}{-s_{i,i+1}}\right)^\e
+ \frac{\zeta_2}{2} + E^{(1)}(u,v,w),
\label{M1loop}
\ee
where we have omitted an overall factor of $e^{-\e \gamma_E}/(4\pi)^{-\e}$,
with $\gamma_E$ the Euler-Mascheroni constant, and where 
\begin{align}
E^{(1)}(u,v,w) &= - 2 \big( \Li_2(1-u) + \Li_2(1-v) + \Li_2(1-w) \big)  \nonumber \\
&\qquad \qquad - \ln u \ln v - \ln v \ln w - \ln w\ln u + 4 \zeta_2 
\label{E1}
\end{align}
is the finite, dual conformally invariant part.\footnote{%
We chose the $\zeta_2$ part of $E^{(1)}$ so that $E^{(1)}$ has
no constant under the logarithms in the soft limit $u,v\to0$.
In any event, we will shift to another normalization later.}

For our ``BDS-like'' ansatz here, we (initially) use the minimal
infrared-divergent part of $M_3^{(1)}(\e,s_{ij})$, namely
\be
\hat{M}_3^{(1)}(\e,s_{ij})
= - \frac{1}{\e^2} \sum_{i=1}^3 \left(\frac{\mu^2}{-s_{i,i+1}}\right)^\e
+ \frac{\zeta_2}{2}.
\label{M1hatloop}
\ee
When we convert from BDS normalization~(\ref{remainderdef}) to
BDS-like normalization,
\be
 \mathcal{F}_{3}^{\rm MHV} = \mathcal{F}_3^{\text{BDS-like}} \times E \,,
\label{BDSlikedef}
\ee
$E^{(1)}$ gets exponentiated along with the remainder
function $R_3$ in \eqn{remainderdef}.
That is, the finite (BDS-like normalized) form factor $E$ and
the remainder function\footnote{Henceforth, we drop the $n=3$ subscript.}
$R \equiv R_3$ are related by
\be
E = \exp\biggl[ \frac{1}{4} \Gcusp E^{(1)} + R \biggr] \,,
\label{EtoR}
\ee
where the cusp anomalous dimension is~\cite{Beisert:2006ez}
\be
\frac14 \Gcusp(g^2)\ =\ g^2 - 2\,\zeta_2\,g^4 + 22\,\zeta_4\,g^6
- \Bigl[ 219\, \zeta_6 + 8 \, (\zeta_3)^2 \Bigr] \, g^8 + \cdots.
\label{Gcusp}
\ee

What consequences does this have for the symbol of $E^{(2)}$?
Expanding \eqn{EtoR} to order $g^4$, we have
\be
\mathcal{S}\Big[ E^{(2)} \Big]
= \mathcal{S} \Big[ R^{(2)} + \frac{1}{2} [ E^{(1)} ]^2 \Big] \,.
\label{E2fromR2}
\ee
Notice that in several terms in the symbol~(\ref{R32symbol}),
the letter $(1-u)$ appears adjacent to $(1-v)$ or $(1-w)$.
However, these appearances are all attributable to the term
$\smash{-2\bigl[\sum_i {\rm Li}_2(1-1/u_i) \bigr]^2}$
in \eqn{eq:two_loop_remainder}. 
For example, the symbol of $\mathrm{J}_4$ is
\be
\cS \Big[ \mathrm{J}_4(z) \Big]
= - \frac{1}{2} \bigg[ z \otimes z \otimes z \otimes \frac{z}{(1-z)^2} \bigg] \,,
\ee
which has the letters $(1-u_i)$ only in the fourth entry.
Using ${\rm Li}_2(1-1/u) = - {\rm Li}_2(1-u) - \tfrac{1}{2} \ln^2 u$,
it is easy to see that the terms with $(1-u_i)$ adjacent to
$(1-u_j)$ for $i\neq j$ are all cancelled by adding $\frac{1}{2} [ E^{(1)} ]^2$
to the remainder function.

Thus, the symbol of $E^{(2)}$ obeys the novel ES-like restrictions
\begin{align} 
\cancel{\ldots \otimes 1-u\otimes 1-v \otimes \ldots},
\label{eq:ExtSteinmannomuomv}
\end{align}
plus the five other conditions generated by the $S_3$
dihedral symmetry \eqref{eq:dihedral}.\footnote{%
  The same adjacency condition was also observed in
  ref.~\cite{Chicherin:2020umh},
  which appeared on the arXiv on the same day as this paper.}
Using compatibility graphs, the same restriction can be seen to hold to
all loop orders for planar ladder-box integrals~\cite{Panzer:2015ida},
including the explicit three-loop result of ref.~\cite{DiVita:2014pza}.
As we will discuss further in section~\ref{sec:generalintegrals},
there is considerable evidence at two loops
that these restrictions are broadly applicable to all processes with
the same kinematics,
i.e.~four-point scattering amplitudes with one massive external leg
and three massless ones, and all massless internal lines,
planar \emph{and} non-planar.
Correspondingly, we will adopt \eqn{eq:ExtSteinmannomuomv}
as part of the definition of the {\it form factor function space}
$\cM$ in the next section.

The condition~(\ref{eq:ExtSteinmannomuomv}) does not appear to arise
from any standard (extended) Steinmann relation~\cite{Steinmann,Steinmann2};
in fact, the letters $1-u_i$ are not associated with physical thresholds.
On the other hand, we understand these conditions in the pentabox ladder
integrals~\cite{Drummond:2010cz,Caron-Huot:2018dsv}.
As will be discussed further in appendix~\ref{app:pentaladders},
these integrals belong to both $\cM$
and the space of heptagon functions relevant for seven-point amplitudes.
As heptagon functions, they inherit adjacency
restrictions from (extended) Steinmann relations (or cluster
adjacency conditions) governing this
space~\cite{Dixon:2016nkn,Drummond:2017ssj}, and these restrictions
imply \eqn{eq:ExtSteinmannomuomv}.

\subsection{Kinematic regions}
\label{sec:kinematicregions}

\begin{figure}[t]
\begin{center}
\includegraphics[width=5.5in]{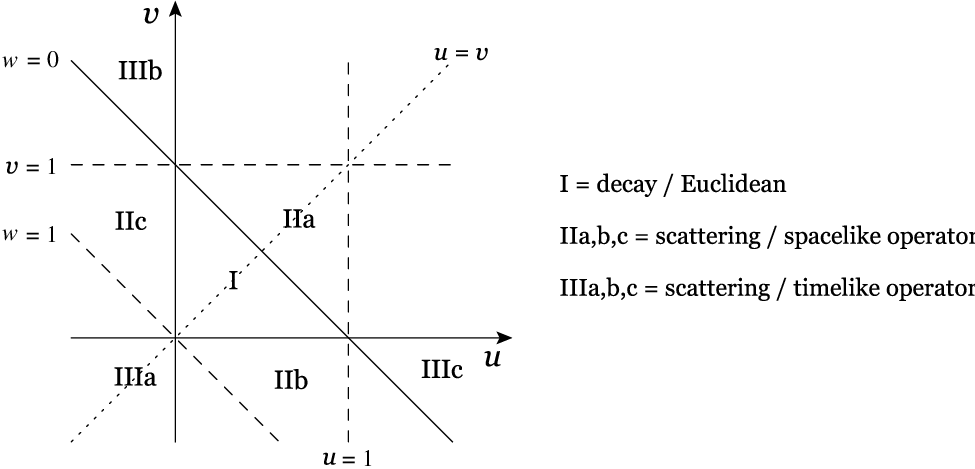} 
\end{center}
\caption{Kinematics of the three-point form factor in the $(u,v)$ plane,
  where $w=1-u-v$.  The Euclidean region is the triangle $0 < u,v,w < 1$,
  and the solid edges bounding it correspond to the near-collinear limits
  where the FFOPE data can be accessed. The other physical regions are
  described further in the text.}
\label{Fig:uvplane}
\end{figure}

Finally, let us discuss the various kinematic regions that can be
accessed.  All real values of $(u,v)$ correspond to physical scattering
or decay processes, as depicted in \Fig{Fig:uvplane}.
In the Euclidean region I, where all four of the Mandelstam invariants
are negative and $0 < u, v, w < 1$, the form factor is manifestly real.
For infrared-finite expressions, this region is equivalent to the
pseudo-Euclidean region with all four invariants positive, which
describes the decay of the (time-like) operator insertion
into three massless particles --- for instance, a
Higgs boson decaying into three positive-helicity gluons, $H \to g^+ g^+ g^+$.
Scattering region IIa, where $w < 0 < u,v$, describes instead a scattering
process involving a space-like operator (similar to deep-inelastic scattering),
such as a (space-like) Higgs boson and a gluon scattering
into two gluons, $H g^- \to g^+  g^+$. 
Scattering regions IIb and IIc are obtained by cyclically permuting $(u,v,w)$
from region IIa.  Finally, scattering region IIIa with $u,v < 0 < w$,
as well as its images under cyclic permutations, IIIb and IIIc,
describe time-like Higgs production, say $g^- g^- \to H g^+$.

The dashed lines in \Fig{Fig:uvplane} correspond to potential
spurious poles, because they coincide with the vanishing loci of the letters $1-u$, $1-v$,
and $1-w$.  On the line $u=1$, the letters \eqref{alphabet2} of $\mathcal{S}_3$ become $\{v,1-v,1+v\}$,
and so the functions in $\cM$ all collapse to HPLs $H_{\vec{a}}(v)$ with $a_i \in \{0,1,-1\}$,
making it straightforward to plot the form factor or remainder
function there.
On the dotted line with $u=v$, the letters become $\{u,1-u,1-2u\}$,
which maps to the same function space with argument $2u-1$.
In section \ref{sec:linevalues}, we will plot the remainder function on the dashed line $u=1$ and
the dotted line $u=v$ through five loops.

\section{The Form Factor Function Space}
\label{sec:func_space}

The polylogarithms that contribute to the one- and two-loop form factor 
have a number of notable features, which can be abstracted away from 
the specific weight-two and weight-four functions $E^{(1)}$ and $E^{(2)}$. 
In particular, the polylogarithms that appear in these functions can be 
individually chosen to obey certain branch cut conditions and restrictions 
on their adjacent symbol letters. In combination with the symbol 
alphabet~\eqref{alphabet}, we generalize these properties to all transcendental 
weights to define the three-point form factor space of functions $\cM$. 
Specifically, we define $\cM$ to be the space of polylogarithms that satisfies 
the following criteria:
\begin{enumerate}
\item[(i)] {\bf Symbol Alphabet:} their symbol only involves letters that
  appear in $\mathcal{S}_3$, as given in \eqn{alphabet} or equivalently \eqn{alphabet2},
\item[(ii)] {\bf Branch Cut Condition:} they develop logarithmic branch cuts
  only at physical thresholds, corresponding to their symbol's first entries
  belonging to $\{u,v,w\}$,
\item[(iii)] {\bf Extended-Steinmann-Like:} the letter $1-u$ never appears
  adjacent to $1-v$ or $1-w$ in the symbol, nor $1-v$ next to $1-w$.
\end{enumerate}
We conjecture that this space of functions contains the perturbative
three-point MHV form factor to all loop orders. 

Since $\cM$ contains only polylogarithms, it is graded by
transcendental weight. Namely, we can decompose
\be
\cM = \bigoplus_{w=0}^\infty \cM_w\, ,
\ee
where $\cM_w$ is the space of polylogarithms of weight $w$ that
satisfy the above constraints. It is analogous to the hexagon and heptagon
function spaces relevant to six- and seven-particle scattering in planar
$\mathcal{N}=4$ sYM theory~\cite{Dixon:2013eka,Dixon:2014iba,Dixon:2014voa,%
  Drummond:2014ffa,Dixon:2015iva,Caron-Huot:2016owq,Dixon:2016apl,%
  Dixon:2016nkn,Drummond:2018caf,Caron-Huot:2019vjl,%
  Caron-Huot:2019bsq},
and it can be constructed iteratively in the weight.
We also assume that the $L$-loop form factor has weight $2L$,
i.e.~$E^{(L)} \in \cM_{2L}$.   We first describe how this
iterative construction can be carried out using the coproduct formalism,
and then we discuss how the dimension of $\cM_w$ grows with
the weight $w$.

\subsection{Construction of the space \texorpdfstring{$\cM$}{M}}

The method of building polylogarithmic spaces of functions from a fixed
symbol alphabet has been described in a number of places in the literature
(see for instance refs.~\cite{Dixon:2013eka,Caron-Huot:2020bkp} or appendix D
of ref.~\cite{Dixon:2015iva}), so we here describe it only briefly.
The basic idea is to build the space of coproducts that correspond to
the desired polylogarithms, rather than the functions themselves.
This simplifies the description of the function space at each weight,
at the cost of making certain properties of the functions non-manifest.
Information about integration constants also has to be retained separately.

To begin our iterative construction of $\cM$, we first determine
which logarithms exist in $\cM_1$. From condition~(i), we have a
candidate six-dimensional space of logarithms. However, as described by
condition~(ii), form factors are only expected to have discontinuities
at thresholds for physical particle production, i.e.~where (multiple)
internal propagators in Feynman integrals can go on shell.
Since we are in a massless theory, this can only happen when one of the
Mandelstam invariants vanishes. From equation~\eqref{uvwdef}, it is clear
that setting any single Mandelstam invariant to zero causes $u$, $v$, or $w$
to vanish (or all of them to become infinite); in no case does it cause any
of them to approach 1. Thus, the only functions appearing in $\cM$ at weight
one are $\ln u$, $\ln v$, and $\ln w$.

To proceed to higher weights, we construct the space of coproducts
corresponding to the functions in $\cM_w$, rather than the
functions themselves. More specifically, at each weight we start from
an ansatz for the coproduct component involving functions of one lower
weight in the first entry:
\be
\Delta_{w-1,1} \cM_w  \subset
\sum_{i,j} c_{ij} F^{(i)}_{w-1} \otimes \ln \phi_j \, ,
\label{eq:ansatz_double_coprod_A}
\ee
where $F^{(i)}_{w-1}$ is the $i^{\text{th}}$ function in $\cM_{w-1}$,
$\phi_j$ is the $j^{\text{th}}$ symbol letter in $\mathcal{S}_3$, and the
$c_{ij} \in \mathbb{Q}$ are undetermined coefficients. The coproduct of
each function in $\cM_w$ will be given by some value of the
coefficients $c_{ij}$ in this ansatz, but not all values of these
coefficients correspond to valid functions in $\cM_w$.
Thus, we need to constrain this ansatz further.  

By assumption, each of the functions $F^{(i)}_{w-1}$ satisfy
constraints (i)-(iii). Thus, condition~(i) is automatically satisfied
by our ansatz for $\cM_w$ since each of the $\phi_j$ are also
drawn from $\mathcal{S}_3$. To impose condition~(iii), we replace each
function $F^{(i)}_{w-1}$ by its coproduct component
$\smash{\Delta_{w-2,1} F^{(i)}_{w-1}}$. This gives us the double coproduct 
\be
\sum_{i,j} c_{ij}\big( \Delta_{w-2,1}F^{(i)}_{w-1} \big) \otimes \ln \phi_j
=  \sum_{i,j,k} c_{ij} \big(F^{(i)}_{w-2}\big)^{\phi_k} \otimes
\ln \phi_k \otimes \ln \phi_j  \,,
\label{eq:ansatz_double_coprod_B}
\ee 
where we have introduced notation in which we denote the
`$\phi_k$ coproduct entry' of a function $F$ with a superscript,
as $F^{\phi_k}$. The ES-like conditions can then be imposed on
the double-coproduct ansatz~\eqref{eq:ansatz_double_coprod_B} by requiring
that
\be
F^{1-u,1-v} = 0\, , \label{ESint}
\ee
plus all dihedral permutations, where $F^{\phi_i,\phi_j}$ denotes the
linear combination of functions appearing in the first entry of the
tensor product~\eqref{eq:ansatz_double_coprod_A} with second and third
entries $\ln \phi_i$ and $\ln \phi_j$.

In addition to this constraint, we must also require that $F$ is a
genuine function. We can
ensure this by solving the integrability conditions on adjacent pairs of
symbol letters. Taking into account the conditions~\eqref{ESint} that we
have already imposed, it is sufficient to require
\be
F^{u,v} - F^{v,u} = F^{v,1-w} - F^{1-w,v},
\label{intonly}
\ee
plus all dihedral permutations. Together, equations~\eqref{ESint}
and~\eqref{intonly} generate 12 conditions on our ansatz.

Finally, we must impose the branch cut condition (ii) on our ansatz.
This can be done by requiring
\be
F^{1-u}(1,v,w)\Big|_{v,w\to0} = 0,
\label{branchcutcondition}
\ee
plus the two cyclically related conditions
$F^{1-v}(u,1,w)|_{w,u\to0} = F^{1-w}(u,v,1)|_{u,v\to0} = 0$.
These conditions prevent the logarithms $\ln(1-u)$, $\ln(1-v)$, $\ln(1-w)$
from appearing at higher weights when multiplied by zeta values.
As discussed above, such functions have branch cuts in unphysical locations,
so they must be forbidden.
In principle, \eqn{branchcutcondition} could be imposed at other points
on the line $u=1$.  However, it is simplest to impose the condition
here, because functions in $\cM$ in the vicinity
of $u=1$ with $v,w \to0$ are simple polynomials in $\ln v$ and $\ln w$,
with zeta-valued coefficients.  We require the vanishing of these polynomials
for any $F^{1-u}$.  That is, $F^{1-u}$ should vanish as $(u,v) \to (1,0)$ from
any angle in the $(u,v)$ plane.  This constraint removes functions even
at symbol level, unlike what happens for the hexagon or heptagon functions.
We can also fix the constant of integration for an arbitrary function
$F\in\cM$ at the same point $(1,v\to 0,w \to 0)$; for example, we can require
that the constant in the polynomial in $\ln v$ and $\ln w$
at this point is zero for each function.

\subsection{Growth of the space \texorpdfstring{$\cM$}{M}}

Solving the conditions outlined in the last section at symbol level through
weight eight,
we find exactly $3^w$ independent functions at each weight $w$.
The generating function for this dimensionality, $d(w) = 3^w$, is simply
\be
d_{\cM,\,{\rm symb}}(t) \equiv \sum_{w=1}^\infty d(w) t^w = \frac{1}{1-3t} \,.
\label{countsymbols}
\ee
Such a simple dimensionality begs for an all-orders construction,
but so far we have just let the computer solve the conditions.

If we had not imposed the ES-like relations, we would expect a faster
asymptotic growth rate of $4^w$.  This growth rate stems from the fact that
four of the letters in \eqn{alphabet2} contain $u$.  Hence
$G_{0,\vec{a}}(u)$, $G_{1,\vec{a}}(u)$, $G_{1-v,\vec{a}}(u)$, and
$G_{-v,\vec{a}}(u)$ all provide possible functions at weight $w$,
for each allowed function $G_{\vec{a}}(u)$ at weight $(w-1)$.
At weight eight, we would expect about 10 times as many functions without
the ES-like relations, making bootstrapping much more difficult.

When $v\to0$, the symbol alphabet collapses to $\{u,1-u,v\}$.  Thus,
on the $v\to0$ line, the limiting behavior of the function space $\cM$
just involves logarithms in $v$, and HPLs $H_{\vec{a}}(1-u)$
with $a_i \in \{0,1\}$~\cite{Remiddi:1999ew}.
The constants associated with these functions are MZVs,
which (motivically) have the generating
function~\cite{Zagier:1994,Broadhurst:1996kc,Brown:2011ik}
\be
d_{\rm MZV}(t) \equiv \sum_{w=0}^\infty d_{\rm MZV}(w) t^w
= \frac{1}{1 - t^2 - t^3}
= 1 + t^2 + t^3 + t^4 + 2 t^5 + 2 t^6 + \ldots \,
\label{eq:MZVindep}
\ee
We define $\cM$ to include constant functions for all such MZVs.
Thus the dimensionality of the full function space $\cM$ is generated by
\bea
d_{\rm MZV}(t) \times d_{\cM,\,{\rm symb}}(t)
&=& \frac{1}{1 - t^2 - t^3} \, \frac{1}{1-3t} \label{countfunctions}\\
&=& 1 + 3 t + 10 t^2 + 31 t^3 + 94 t^4 + 284 t^5 + 854 t^6 + 2565 t^7
+ 7699 t^8 + \cdots
\nonumber
\eea
In practice, we have constructed the full function space through
weight eight.

Many of the functions in $\cM$ are quite simple.  Any polynomial
in $\ln u, \ln v, \ln w$ is in the space. Furthermore, if a function $F$ is
in the space, then so is $\ln u_i \times F$, where $u_1 = u$, $u_2=v$, $u_3=w$.
Also in $\cM$ are the functions
$H_{\vec{a}}(1-u_i)$ with $a_k \in \{0,1\}$, where the last $a_k$ must be 1.
On the other hand, we cannot multiply two of these functions for
different $u_i$ together.  For example, the product
$H_{0,1}(1-u) H_{0,1}(1-v)$ is not in $\cM$,
because its symbol is the shuffle of
$u\otimes(1-u)$ with $v\otimes(1-v)$, which contains terms where
$(1-u)$ and $(1-v)$ are adjacent, which is not allowed.

Let us define the ``simple'' functions in $\cM$
to be all of the functions $H_{\vec{a}}(1-u_i)$
multiplied by arbitrary polynomials in all of the $\ln u_j$.
The generating function for this space of simple functions is
\be
d_{\rm simple}(t) = \frac{1}{(1-t)^3}
\biggl[ - 2 + \frac{3(1-t)^2}{1-2t} \biggr] \,,
\label{countsimple}
\ee
where the factor of $1/(1-t)^3$ accounts for the polynomials in all
the $\ln u_j$.  The remaining functions depend irreducibly on two
variables, i.e.~they are true 2dHPLs~\cite{Gehrmann:2000zt}.
We do not yet have a closed-form construction of such functions in $\cM$.
However, we can count them.
We remove the ones that are simply powers of logs times lower weight
2dHPL functions by multiplying by $(1-t)^3$.
Thus the generating function for the 2d functions is
\bea
d_{\rm 2d}(t)
&=& (1-t)^3 \biggl[ d_{\cM,\,{\rm symb}}(t) - d_{\rm simple}(t) \biggr]
= \frac{2t^3(1+t)}{(1-2t)(1-3t)}\nonumber\\
&=& 2 t^3 + 12 t^4 + 48 t^5 + 168 t^6 + 552 t^7 + 1752 t^8 + 5448 t^9
+ 16728 t^{10} + \cdots~~~~~
\label{count2d}
\eea
at symbol level.

Non-classical polylogarithms can first appear at weight four.
Applying the Lie cobracket test in ref.~\cite{Goncharov:2010jf},
we find that there are 9 non-classical functions, which are contained within
the set of 12 weight-four 2dHPLs in \eqn{count2d}. For example,
the function $r_i^{(2)}$ appearing in the operator form factor studied
in ref.~\cite{Brandhuber:2014ica} contains non-classical polylogarithms
and appears in the space $\cM$.


\section{Bootstrapping the Three-Point Form Factor}
\label{sec:bootstrapping}

Having constructed $\cM$ up to weight eight, we now want to identify the
planar ${\cal N}=4$ sYM form factor within this space. We first assume that the
$L$-loop contribution to the form factor has uniform transcendental
weight $2L$, as is true for scattering amplitudes in this theory.
A number of additional constraints were used to bootstrap the two-loop
remainder function in ref.~\cite{Brandhuber:2012vm}. However, as discussed in 
section~\ref{sec:background}, the remainder function does {\it not} 
belong to $\cM$ because it does not satisfy the
constraint~(\ref{eq:ExtSteinmannomuomv}). 
Only the form factor $E$, defined via the relation~\eqref{EtoR}, 
satisfies this constraint. 

In fact, there is an even more optimal normalization for bootstrapping the three-point form factor. Its definition just differs from that of $E$ by an exponentiated product of logs:
\be
\EE = \exp\biggl[ \frac{1}{4} \Gcusp \EE^{(1)} + R \biggr] \,,
\label{EEtoR}
\ee
where
\bea
\EE^{(1)}(u,v,w)
&=& E^{(1)} - \ln^2 u - \ln^2 v - \ln^2 w
+ \ln u \ln v + \ln v \ln w
+ \ln w \ln u - 4 \zeta_2 \nonumber\\
&=& - 2 [ \Li_2(1-u) + \Li_2(1-v) + \Li_2(1-w) ]
- \ln^2 u - \ln^2 v - \ln^2 w \nonumber\\
&=& 2 \biggl[ \Li_2\biggl(1-\frac{1}{u}\biggr)
            + \Li_2\biggl(1-\frac{1}{v}\biggr)
            + \Li_2\biggl(1-\frac{1}{w}\biggr) \biggr] \,.
\label{EE1}
\eea
The shift from $E$ to $\EE$ corresponds to using a different BDS-like
normalization of the form factor.
Because $\EE^{(1)} - E^{(1)}$ is polynomial in $\{ \ln u, \ln v, \ln w \}$,
this shift does not spoil the ES-like relations obeyed by $E$.
However, $\EE^{(1)}$ has a non-zero double
discontinuity in $u$, unlike $E^{(1)}$.
Even so, at high loop orders, this disadvantage
is greatly outweighed by three advantages of $\EE^{(1)}$:
\begin{enumerate}
\item[(a)] it satisfies the same final-entry condition as the remainder function,
   $\EE^{(1),\,u} = - \EE^{(1),\,1-u}$,
  \item[(b)] ${\cal E}$ satisfies satisfies simplified multiple-final-entry conditions,  
  \item[(c)] the space of functions contained in all the iterated coproducts
    of $\EE$ is much smaller than that for $E$.
\end{enumerate}
Hence, although our bootstrap was initially carried out in terms of the form factor function 
$E$, we will also present the bootstrap constraints on $\EE$. In particular, we will
show that $\EE$ can be bootstrapped through four-loop order using very little information
from the FFOPE.

Before considering the constraints on $E$ or $\EE$, let us first summarize the constraints that can be imposed on the remainder function:
\begin{enumerate}
\item {\bf Dihedral symmetry}:  $R(u,v,w)$ is invariant under any
permutation of $u,v,w$.
\item {\bf Final entry}: The remainder function obeys $R^{u_i}+R^{1-u_i} = 0$,
  resulting in only three linearly independent final entries.
\item {\bf Next-to-final entry}:  There are only six independent
next-to-final entries, to be described below.
\item {\bf Collinear limit}: The leading-power collinear limit of $R$
should vanish.
\item {\bf Discontinuity}: The $L^{\rm th}$ discontinuity in $u$ of $R^{(L)}$
should vanish.\footnote{This is the $L$-loop generalization of the second-entry constraint utilized in ref.~\cite{Brandhuber:2012vm} to bootstrap two loops, and it follows from general properties of the OPE~\cite{Gaiotto:2011dt}.}
\item {\bf Near-collinear limit}:  The near collinear limit of $R$
  should agree with the predictions of the
  FFOPE~\cite{Sever:2020jjx,Toappear1,Toappear2}.  
\end{enumerate}
Some of these constraints, such as the restriction on next-to-final entries,
are new results of our analysis through five loops. We will discuss these
constraints in detail below. 

The translation of some of these constraints on $R$ into constraints
on $E$ or $\EE$ is not entirely simple. Consider for instance the
vanishing of the $L^{\rm th}$ discontinuity of $R^{(L)}$ in the $u$ channel. 
From \eqn{E1}, we have that
\bea
{\rm Disc}_u E^{(1)} &=& 2 \ln(1-u) - \ln v - \ln w,
\label{discE1}\\
\bigl[ {\rm Disc}_u \bigr]^{2} \, E^{(1)} &=& 0 \,.
\label{ddiscE1}
\eea
Thus, if we take the coefficient of $g^{2L}$ in \eqn{EtoR}
and compute its $L^{\rm th}$
discontinuity, we can immediately impose the condition that the 
second discontinuity of $E^{(1)}$ vanishes and similarly that 
the $L^{\rm th}$ discontinuity of $R^{(L)}$ vanishes for all $L$. 
Doing so, we see that only the
pure $E^{(1)}$ term in \eqn{EtoR} can contribute at symbol level. 
Using the Leibniz rule, we furthermore have that
\be
{\rm Disc}_u \Bigl\{ [{\rm Disc}_u E^{(1)}]^k \exp[ g^2 \, E^{(1)} ] \Bigr\}
= g^2 [{\rm Disc}_u E^{(1)}]^{k+1} \exp[ g^2 \, E^{(1)} ] \,.
\label{discexp}
\ee
Thus, the vanishing discontinuity condition on the remainder function
corresponds to the requirement that
\be
[ {\rm Disc}_u ]^L E^{(L)} = [ {\rm Disc}_u E^{(1)} ]^L \,
\label{discEL}
\ee
at the level of the symbol.

Similarly, we can work out the final-entry condition on $E$.
Using the fact that $R^u + R^{1-u} = 0$ and that
\be
E^{(1),\,u} + E^{(1),\,1-u} = 2 \ln u - \ln v - \ln w \equiv X \,,
\label{finalentryE1}
\ee
one derives the following conditions at higher loops:
\bea
E^{(2),\,u} + E^{(2),\,1-u} &=&
[ E^{(1)} - 2 \zeta_2 ] \, X \,,
\label{finalentryE2}\\
E^{(3),\,u} + E^{(3),\,1-u} &=&
[ E^{(2)} - 2 \zeta_2 E^{(1)} + 22 \zeta_4 ] \, X \,,
\label{finalentryE3}\\
E^{(4),\,u} + E^{(4),\,1-u} &=&
[ E^{(3)} - 2 \zeta_2 E^{(2)} + 22 \zeta_4 E^{(1)}
 - 219 \zeta_6 - 8 (\zeta_3)^2 ] \, X \,,
\label{finalentryE4}\\
E^{(5),\,u} + E^{(5),\,1-u} &=&
\biggl[ E^{(4)} - 2 \zeta_2 E^{(3)} + 22 \zeta_4 E^{(2)}
 - ( 219 \zeta_6 + 8 (\zeta_3)^2 ) E^{(2)} \nonumber\\
&&\hskip0cm\null
 +\frac{7096}{3} \zeta_8 + 32 \zeta_2 (\zeta_3)^2
 + 160 \zeta_3 \zeta_5 \biggr] \, X \,.
\label{finalentryE5}
\eea
These relations hold at full function level.
In practice it requires some work to utilize them, because the
multiplication of the logarithms in $X$ by the lower-loop form factors
is not well-supported by the $\{n-1,1\}$ coproduct organization
of the function space in the bulk.  However, one can make use of
the behavior of each function in the collinear and near-collinear limits,
as well as on the dotted and dashed lines in \Fig{Fig:uvplane}, 
to pin down where the functions on the right-hand sides of
these equations sit within the function space.

In the $\EE$ normalization, the final-entry condition
can be phrased much more simply. Namely, because $\EE^{(1),\,u} + \EE^{(1),\,1-u} = 0$,
we have
\be
\EE^{(L),\,u} + \EE^{(L),\,1-u} = 0,
\label{EEfinalentry}
\ee
to all loop orders $L$. This normalization also turns out to obey simple multiple-final-entry conditions, which we will present in section~\ref{sec:mult_final_entries}.

\subsection{Near-collinear limit via integrability}
\label{sec:FFOPE}

An important source of data for the perturbative form factor bootstrap
stems from the near-collinear limit.
Using the recently developed form factor operator product expansion, or FFOPE,
the form factor can be determined in an expansion around the collinear limit
for any value of the planar coupling constant \cite{Sever:2020jjx,Toappear1,Toappear2}.
Let us briefly review this construction.

The FFOPE is based on the dual description of the form factor in terms
of a periodic Wilson loop \cite{Alday:2007he,Maldacena:2010kp,
  Brandhuber:2010ad,Ben-Israel:2018ckc,Bianchi:2018rrj}.
Similar to scattering amplitudes, this dual Wilson loop is defined via dual
points $x_i$, where $x_{i+1}-x_i=p_i$. However, since $\sum_ip_i=q$,
the dual Wilson loop for form factors is periodic instead of closed:
$x_{i+n}-x_i=q$.
Note that dual conformal symmetry acts on both the dual momenta and the periodicity
constraint.
Using dual momentum space, the expectation value of the
(suitably regularized) Wilson loop can be written in terms of
dual conformal cross ratios \cite{Sever:2020jjx}.
For the three-point form factor, we have
\begin{align}
  \frac{(x_1-x_3)^2(x_4-x_6)^2}{(x_1-x_4)^2(x_3-x_6)^2}
  &= \frac{1}{(1+e^{-2\tau}+e^{2\sigma})^2} = u^2\,,
\\
 \frac{(x_2-x_4)^2(x_5-x_7)^2}{(x_2-x_5)^2(x_3-x_7)^2}
 &= \frac{1}{(1+e^{2\tau})^2}=v^2\,.
\end{align}
We can also express the cross ratios in terms of 
$T = e^{-\tau}$ and $S = e^\sigma$,
in terms of which they are given by~\cite{Sever:2020jjx}
\be
u = \frac{1}{1 + S^2 + T^2} \,, \quad
v = \frac{T^2}{1 + T^2} \,, \quad 
w = \frac{1}{(1 + T^2)(1 + S^{-2} (1 + T^2))} \,.
\label{uvwtoST}
\ee
Note that, in principle, algebraic symbol letters could appear in the form
factor at higher weight;
however, the fact that the OPE expansion parameter is $T^2$ rather
than $T$~\cite{Sever:2020jjx} provides strong motivation to consider only the six letters in
\eqn{alphabet}.\footnote{In higher-multiplicity form factors, it is also reasonable to expect elliptic polylogarithms and integrals over higher-dimensional manifolds to occur; see for instance refs.~\cite{Laporta:2004rb,MullerStach:2012az,brown2011multiple,Bloch:2013tra,Adams:2013kgc,Adams:2015gva,Adams:2016xah,Bourjaily:2017bsb,Broedel:2017kkb,Adams:2018kez,Bourjaily:2018ycu,Bourjaily:2018yfy,Bourjaily:2019hmc,Bogner:2019lfa,Broedel:2019kmn}.}

The FFOPE immediately applies to a finite ratio of Wilson loops,
$\mathcal{W}_n$, which is defined such that the ultraviolet divergences occurring
at the cusps vanish; see ref.\ \cite{Sever:2020jjx} for details.
It can be translated to the remainder function $R_n$ via 
\begin{equation}
\label{eq:relation-W-R}
  {\cal W}_n
  =\exp\left[\frac{1}{4}\Gcusp\mathcal{W}^{(1)}_{n}+R_n\right]\,,
\end{equation}
where 
\begin{equation}
  \mathcal{W}_{3}^{(1)}
  = 4\sigma^2 - 2 \Li_2(- e^{-2\tau}) + 2\Li_2(-e^{-2\tau} - e^{2\sigma})
  + 2 \Li_2(- e^{-2\tau} - e^{-2\sigma}(1 + e^{-2\tau})^2)+\frac{\pi^2}{3}\,,
\end{equation}
 and $\Gamma_{\text{cusp}}$ is the cusp anomalous dimension \eqref{Gcusp}.
 
The FFOPE is the large-$\tau$ expansion of the ratio $\mathcal{W}_{n}$,
and is given for $n=3$ by~\cite{Sever:2020jjx}
\begin{equation}
  \mathcal{W}_{3}=\int\!\!\!\!\!\!\!\!\!\sum_{\psi}e^{-E_\psi \tau+ip_\psi\sigma}\mathcal{P}(0|\psi)\mathcal{F}(\psi)\,,
\end{equation}
where the sum is over a basis of eigenstates $\psi$ of the
Gubser-Klebanov-Polyakov (GKP) flux tube. All ingredients in this
expansion are known at finite coupling: the energies $E_\psi$ and
momenta $p_\psi$ of states were found in ref.\ \cite{Basso:2010in},
the pentagon transitions $\mathcal{P}$ in refs.\ \cite{%
  Basso:2013vsa, Basso:2013aha,Basso:2014koa,Basso:2014nra,Basso:2014hfa,%
  Basso:2015rta,Basso:2015uxa,Belitsky:2014sla,Belitsky:2014lta,%
  Belitsky:2016vyq}
and the form factor transitions $\mathcal{F}$ in refs.\
\cite{Sever:2020jjx,Toappear1,Toappear2}.
At any loop order in the weak-coupling expansion, only flux-tube states
$\psi$ with a finite number of effective excitations contribute.
The integral over the momenta of these excitations can then be done
via residues to obtain a series expansion in $e^\sigma$ to any desired order.

We have used the above procedure to obtain a series expansion of the
terms of order $T^2$ up to $\cO(S^{300})$ through five loops,
but this can be extended easily to higher loop orders and higher
orders in $S$.\footnote{In practice, information from the perturbative
  form factor bootstrap was initially used to determine the higher-order behavior
  of the form factor transition, which in turn provided more subleading
  logarithms of $T$ in the $T^2$ terms in the near-collinear limit,
  which were in turn needed to fix the form factor at higher loop orders.}
We have also generated selected results for the terms of order $T^4$
up to five-loop order.  After a translation to the remainder
function \eqref{eq:relation-W-R}, they agree perfectly with the expansion
of our result for the remainder function.
In appendix~\ref{app:OPEformulas}, we give the results for the
expansion of $R$ in closed form in $S$ at two and partially at three loops;
the higher-loop results are provided in the ancillary files
{\tt T2terms.txt} and {\tt T4terms.txt}.


\subsection{Multiple-final-entry conditions}
\label{sec:mult_final_entries}

One of our initial assumptions is that the remainder function
obeys the same final-entry condition as the six-point amplitude,
${\cal R}^{u_i} + {\cal R}^{1-u_i} = 0$~\cite{CaronHuot:2011kk}.\footnote{%
It might be possible to derive this relation using similar methods
as were used to derive the $\bar{Q}$ equation for closed polygonal
Wilson loops.} As described above, the $L$-loop form factor $E^{(L)}$
does not obey the same relation, since the one-loop form factor $E^{(1)}$ does not respect 
the same final-entry condition. However, the violation of this 
final-entry condition is predictable, as detailed in
eqs.~(\ref{finalentryE2})--(\ref{finalentryE5}).
From \eqn{finalentryE1}, the sum over all three cyclic images
of $E^{(1),\,u} + E^{(1),\,1-u}$ vanishes, and therefore $E^{(L)}$ does
obey one homogeneous final-entry condition,
\be
E^{(L),\,u} + E^{(L),\,v} + E^{(L),\,w}
+ E^{(L),\,1-u} + E^{(L),\,1-v} + E^{(L),\,1-w} = 0.
\label{totsum}
\ee
Thus, $E^{(L)}$ seems to have five final entries. However, we know that only 
three of them can be truly new,
due to the inhomogeneous
relations~(\ref{finalentryE2})--(\ref{finalentryE5}) and the fact that we know
the lower-loop form factors.

\renewcommand{\arraystretch}{1.25}
\begin{table}[!t]
\centering
\begin{tabular}[t]{l c c c c c c c c c c c}
\hline\hline
weight $n$
& 0 & 1 & 2 & 3 & 4 &  5 &  6 &  7 &  8 &  9 & 10
\\\hline\hline
$L=1$
& \green{1} & \green{3} & \blue{1} &  &  &  &  &  &  &  &
\\\hline
$L=2$
& \green{1} & \green{3} & \green{10} & \blue{5} & \blue{1} &  &  &  &  &  &
\\\hline
$L=3$
& \green{1} & \green{3} & \green{10} & 27
& \blue{15} & \blue{5} & \blue{1} &  &  &  &
\\\hline
$L=4$
& \green{1} & \green{3} & \green{10} & \green{31} & 69 & \blue{37}
& \blue{15} & \blue{5} & \blue{1} &  & 
\\\hline
$L=5$
& \green{1} & \green{3} & \green{10} & \green{31} & 88 & 162
& 83 & \blue{37} & \blue{15} & \blue{5} & \blue{1}
\\\hline\hline
\end{tabular}
\caption{The number of independent $\{n,1,1,\ldots,1\}$ coproducts
of the form factor $E^{(L)}$ through $L=5$ loops. A green number
denotes saturation of the space $\cM$ constructed from the bottom up.
A blue number denotes saturation of the space of coproducts, as indicated
by the $(L,n)$ number being the same as the $(L+1,n+2)$ number.}
\label{tab:Ecopdim}
\end{table}

\renewcommand{\arraystretch}{1.25}
\begin{table}[!t]
\centering
\begin{tabular}[t]{l c c c c c c c c c c c}
\hline\hline
weight $n$
& 0 & 1 & 2 & 3 & 4 &  5 &  6 &  7 &  8 &  9 & 10
\\\hline\hline
$L=1$
& \green{1} & \blue{3} & \blue{1} &  &  &  &  &  &  &  &
\\\hline
$L=2$
& \green{1} & \green{3} & \blue{6} & \blue{3} & \blue{1} &  &  &  &  &  &
\\\hline
$L=3$
& \green{1} & \green{3} & \green{9} & \blue{12} & \blue{6} & \blue{3}
& \blue{1} &  &  &  &
\\\hline
$L=4$
& \green{1} & \green{3} & \green{9} & \green{21} & \blue{24} & \blue{12}
& \blue{6} & \blue{3} & \blue{1} &  & 
\\\hline
$L=5$
& \green{1} & \green{3} & \green{9} & \green{21} &  47 & 45
& \blue{24} & \blue{12} & \blue{6} & \blue{3} & \blue{1}
\\\hline\hline
\end{tabular}
\caption{The same as Table~\ref{tab:Ecopdim}, but for $\EE^{(L)}$
  instead of $E^{(L)}$.}
\label{tab:EEcopdim}
\end{table}

Similarly, we can ask how many independent $k^{\rm th}$ final entries
there are, or $\{2L-k,1,\ldots,1\}$ coproducts of the $L$-loop form factor
$E^{(L)}$.  This set of dimensions is given in Table~\ref{tab:Ecopdim}.
In Table~\ref{tab:EEcopdim} we present the same dimensions for $\EE^{(L)}$.
The number of independent functions appearing in the coproduct components of $\EE^{(L)}$ 
is considerably smaller, principally because $\EE^{(1)}$ obeys
the same final-entry condition as $R$.

Since the multiple-final-entry analysis is simpler for $\EE$,
we focus on this normalization.
For $k=1$, the independent final entries can be taken to be
\be
\{ \EE^{1-u},\ \EE^{1-v},\ \EE^{1-w}\} \, .
\label{indepFE1}
\ee
The rest are determined by the relation
\be
\EE^u = - \EE^{1-u} \, ,
\label{ufromomu}
\ee
as well as its dihedral images.
For $k=2$, the independent next-to-final entries can be taken to be
\be
\{ \EE^{u,1-u},\ \EE^{1-u,1-u},
 \ \EE^{v,1-v},\ \EE^{1-v,1-v},
 \ \EE^{w,1-w},\ \EE^{1-w,1-w} \} \,.
\label{indepFE2}
\ee
The final-entry pairs ending in one of the $u_i$ are related to those ending
in $1-u_i$ by taking further coproducts of \eqn{ufromomu}.  So we only need
to specify those ending in a $1-u_i$, and of course $E^{1-u_j,1-u_i} = 0$
for $j \neq i$.  That leaves only the $E^{u_j,1-u_i}$.  They are
all given by dihedral images of the relation,
\bea
\EE^{u,1-v} &=& - \EE^{w,1-w} - \EE^{1-w,1-w} + \EE^{u,1-u} + \EE^{1-u,1-u} \,,
\label{maindoublefinalrelation}
\eea
which was first found empirically at three loops.

Similarly, we find that, starting at three loops, there are 12
linearly independent triple final entries for $\EE$, which can be taken to be
\be
\{ \EE^{u,u,1-u},\ \EE^{u,1-u,1-u},\ \EE^{v,u,1-u},\ \EE^{w,u,1-u},\ \dots \} \,,
\label{indepFE3}
\ee
where the ellipses denote the cyclic images of the first four.
The other triple final entries of the form $\EE^{x,u,1-u}$ and $\EE^{x,1-u,1-u}$
are related to these by
\bea
\EE^{1-u,u,1-u} &=& \EE^{u,1-u,1-u} - \EE^{v,u,1-u} - \EE^{w,u,1-u} \,,
\label{FE3eq1}\\
\EE^{1-v,u,1-u} &=& \frac{1}{3} \Bigl[
  - \EE^{u,v,1-v} + \EE^{u,w,1-w} - \EE^{v,u,1-u} - \EE^{v,v,1-v} + \EE^{v,w,1-w}
\nonumber\\&&\hskip0.3cm\null
- \EE^{v,1-v,1-v} - 2 \, \EE^{w,u,1-u} - \EE^{w,v,1-v}
+ \EE^{w,w,1-w} + \EE^{w,1-w,1-w} \Bigr] \,,
\label{FE3eq2}\\
\EE^{v,1-u,1-u} &=& \frac{1}{3} \Bigl[ - \EE^{v,u,1-u} + \EE^{v,v,1-v}
+ \EE^{v,w,1-w} + \EE^{v,1-v,1-v} + \EE^{w,u,1-u}
\nonumber\\&&\hskip0.3cm\null
- \EE^{w,v,1-v} - \EE^{w,w,1-w} - \EE^{w,1-w,1-w} \Bigr] \,,
\label{FE3eq3}\\
\EE^{1-u,1-u,1-u} &=& \frac{1}{3} \Bigl[  - 2 \, \EE^{u,u,1-u} - 5 \, \EE^{u,1-u,1-u}
+ \EE^{v,u,1-u} + \EE^{v,v,1-v} - \EE^{v,w,1-w}
\nonumber\\&&\hskip0.3cm\null
+ \EE^{v,1-v,1-v} + \EE^{w,u,1-u} - \EE^{w,v,1-v}
+ \EE^{w,w,1-w} + \EE^{w,1-w,1-w} \Bigr] \,,
\label{FE3eq4}\\
\EE^{1-v,1-u,1-u} &=& 0,
\label{FE3eq5}
\eea
as well as the $(v\lr w)$ flip of these relations.
These relations, and \eqn{maindoublefinalrelation}, hold for
all the form factors through five loops.

Finally, Table~\ref{tab:EEcopdim} shows that there are 24 independent
quadruple final entries, starting at four loops.
There is a corresponding set of quadruple final entry relations
that holds through five loops.  We have not yet fully characterized these
relations, or used them as constraints in our bootstrap.

\subsection{Imposing the constraints}

We originally bootstrapped the form factor $E^{(L)}$ through five loops,
and did not make much use of multiple-final-entry relations, relying more on data from the FFOPE~\cite{Sever:2020jjx,Toappear1,Toappear2}.\footnote{An exception to this was at five loops. Since we have not constructed $\cM$ through weight ten, we bootstrapped the five-loop form factor starting from an ansatz for its $\{8,1,1\}$ coproduct, which took into account relations \eqref{maindoublefinalrelation} and \eqref{FE3eq1}--\eqref{FE3eq5}.}
However, the implementation of the constraints for $\EE^{(L)}$ is
generally simpler.  In Table~\ref{tab:Mparameters} we present
the number of parameters remaining as we impose the various
constraints described above on our ansatz for $\EE^{(L)}$,
working in the space of functions $\cM$.
The final-entry condition in that table refers to~\eqn{ufromomu},
while the next-to-final-entry condition refers to~\eqn{maindoublefinalrelation}.
The vanishing of the $L^{\rm th}$ discontinuity of $R^{(L)}$ in $u$
is implemented at symbol level.
At three loops, only one constant has to be fixed using
the OPE data, and it can be done with just the $T^2 \ln^2 T \, S^2 \ln^2 S$
coefficient, while we actually used significantly more OPE data when
we initially constructed $E^{(3)}$.
At four loops, Table~\ref{tab:Mparameters} shows that we need to use the
OPE data down to $T^2 \ln^1 T$ in order to fix all of the 32 parameters
remaining after imposing the other conditions.

\renewcommand{\arraystretch}{1.25}
\begin{table}[!t]
\centering
\begin{tabular}[t]{l c c c }
\hline\hline
$L$                        &  2 &   3 &    4
\\\hline\hline
functions in $\cM$    & 94 & 854 & 7699 \\\hline
dihedral symmetry          & 24 & 176 & 1419 \\\hline
final entry                &  8 &  49 &  354 \\\hline
next-to-final entry        &  5 &  22 &  125 \\\hline
collinear limit            &  0 &   1 &   33 \\\hline
$L^{\rm th}$ discontinuity   &  0 &   1 &   32 \\\hline
OPE $T^2\,\ln^{L-1}T$       &  0 &   0 &   28 \\\hline
OPE $T^2\,\ln^{L-2}T$       &  0 &   0 &   16 \\\hline
OPE $T^2\,\ln^{L-3}T$       &  0 &   0 &   0 \\\hline\hline
\end{tabular}
\caption{Number of parameters left when bootstrapping the form factor
  $\EE^{(L)}$ at $L$-loop order in the function space $\cM$.}
\label{tab:Mparameters}
\end{table}

We do not present the numbers of parameters 
for five loops in Table~\ref{tab:Mparameters}
because redoing the whole construction for $\EE^{(5)}$ would be
too time-consuming.  In bootstrapping $E^{(5)}$, we used the OPE
data all the way down to $T^2\,\ln^{1}T$, as well as the triple-final-entry
conditions, in order to find a unique solution.  The solution was validated
on the $T^2\,\ln^{0}T$ data and selected $T^4\,\ln^{k}T$ data~\cite{Toappear2}.
The fact that the numbers of independent coproducts at weights two, three, and four
in Table~\ref{tab:EEcopdim}
drop significantly when shifting from $E^{(5)}$ to $\EE^{(5)}$
provides additional circumstantial
evidence that the five-loop solution is correct.
We provide the symbols of the form factors $\EE^{(L)}$ through five loops
and the remainder functions $R^{(L)}$ through four loops in the ancillary file
{\tt cEandRsymbols.txt}.

In ref.~\cite{Brandhuber:2012vm} it was observed that the symbol
of the remainder function of the two-loop six-gluon amplitude, ${\cal R}^{(2)}$,
can be written as the sum of two terms. The first term is precisely the symbol
of the three-point form factor $R^{(2)}$. The second term has a final entry
which is the product of the three parity-odd hexagon letters, $y_u y_v y_w$.
This term actually vanishes on the surface $u+v+w=1$,
because $y_u y_v y_w \to -1$ there.  Indeed, it is possible to show
that the two functions coincide on this surface,
${\cal R}^{(2)}(u,v,1-u-v) = R^{(2)}(u,v)$.\footnote{We thank C.~Duhr
  for discussions on this point.}
However, this agreement turns out to be a coincidence that is only true at
two-loop order.  That is, comparing our three-loop remainder symbol
with the corresponding one for the three-loop
six-gluon amplitude~\cite{Dixon:2011pw},
they do not agree on the surface $u+v+w=1$.

\subsection{A smaller function space \texorpdfstring{$\cC$}{C}}

It is clear from Table~\ref{tab:EEcopdim} that there is a smaller
space $\cC \subset \cM$ which contains all of the coproducts
of $\EE$.  Because $\cM$ grows rather rapidly with the weight,
understanding $\cC$ will be key to pushing this form factor
bootstrap beyond five loops.
At weight two, the missing function in $\cC$ is the constant
$\zeta_2$. It gets coupled to the other weight-two functions as follows:
\be
\cC_{w=2} = \biggl\{
\frac{1}{2} \ln^2\biggl(\frac{u_i}{u_{i+1}}\biggr) + \zeta_2 \,,\ \ 
\frac{1}{2} \ln^2\biggl(\frac{u_i u_{i+1}}{u_{i+2}}\biggr) + \zeta_2 \,,\ \ 
H_{0,1}(1-u_i) \biggr\} \,.
\label{Cwt2}
\ee
At weight three, there are 21 functions. They can be sorted into seven
three-orbits under the dihedral symmetry. The constant $\zeta_3$ does
not appear in this space, nor do 6 other symbol-level 
functions that are in $\cM$.
The five simplest three-orbits are
\bea
\cC_{w=3} &\supset& \biggl\{
\frac{1}{6} \ln^3\biggl(\frac{u_i}{u_{i+1}}\biggr)
+ \zeta_2 \, \ln\biggl(\frac{u_i}{u_{i+1}}\biggr) \,,\ \ 
\frac{1}{6} \ln^3\biggl(\frac{u_i u_{i+1}}{u_{i+2}}\biggr)
+ \zeta_2 \, \ln\biggl(\frac{u_i u_{i+1}}{u_{i+2}}\biggr) \,,\ \ 
\nonumber\\&&\hskip0.3cm\null
\frac{1}{6} \ln^3 u_i - H_{0,1,1}(1-u_i) \,, \ \ 
\frac{1}{6} \ln^3 u_i + \ln u_i \, H_{0,1}(1-u_i) + 2 \, H_{0,1,1}(1-u_i) \,,
\nonumber\\&&\hskip0.3cm\null
H_{0,0,1}(1-u_i) + H_{0,1,1}(1-u_i)  \biggr\} \,,
\label{Cwt3}
\eea
but we do not have illuminating representations of the two more
complicated three-orbits.

We have not yet fully characterized $\cC$ at higher weights, but
we can construct a version of it based on the 21
weight-three functions listed in Table~\ref{tab:EEcopdim}.
We construct the weight-four functions by requiring that their $\{3,1\}$
coproducts lie in this 21-dimensional space, as well as imposing the
ES-like, integrability, and branch-cut conditions.
We find 51 symbol-level functions, plus $\zeta_4$.
(It may be possible to discard three of these functions;
the 47 functions in the coproducts of $\EE^{(5)}$ can
be supplemented with 2 more functions from the pentaladder space
$\mathcal{P}_4$ in appendix~\ref{app:pentaladders} to obtain
a space that appears to be complete.  But it is more conservative
to keep the three functions for now.)
Continuing iteratively through weight eight,
we find the dimensions listed in Table~\ref{tab:FvsC},
along with those for $\cM$.  The symbol-level dimensions are
consistent with the generating function
\be
d_{\cC,\,{\rm symb}}(t) = \frac{ 1 - 2 t - 6 t^3 }{(1-2t)(1-3t)} \,,
\label{countCsymbols}
\ee
which also grows asymptotically like $3^w$,
although we emphasize again that the true $\cC$ is likely
to be considerably smaller.

\renewcommand{\arraystretch}{1.25}
\begin{table}[!t]
\centering
\begin{tabular}[t]{l c c c c c c c c c c c}
\hline\hline
weight $n$
& 0 & 1 &  2 &  3 &  4 &   5 &   6 &    7 &    8  \\\hline\hline
functions in $\cM$
& 1 & 3 & 10 & 31 & 94 & 284 & 854 & 2565 & 7699  \\\hline
functions in $\cC$
& 1 & 3 &  9 & 21 & 52 & 134 & 356 &  969 & 2695  \\\hline\hline
\end{tabular}
\caption{Number of functions in $\cM$ and in the trimmed space $\cC$,
 as a function of the weight.}
\label{tab:FvsC}
\end{table}

We have redone the bootstrap construction in the function space
$\cC$ through four loops,
in order to see how much it lessens the requirements for OPE data.
Because we do not have a usable weight-eight basis for $\cC$ yet, at four loops we
worked at the level of the $\{7,1\}$ coproducts, and at symbol level.
Due to this restriction, we impose
dihedral symmetry and the final-entry condition simultaneously,
and we report the numbers of surviving parameters at the level of the symbol.
We see from Table~\ref{tab:Cparameters}
that through four loops there are considerably fewer parameters in early
stages than in Table~\ref{tab:Mparameters},
although similar amounts of OPE information is required at the last stage.
The smaller size of the space $\cC$ should bode well for bootstrapping the
three-point form factor beyond five loops.

\renewcommand{\arraystretch}{1.25}
\begin{table}[!t]
\centering
\begin{tabular}[t]{l c c c }
\hline\hline
$L$                                     &  2 &   3 &          4
\\\hline\hline
symbols in $\cC$                        & 51 & 339 &       2571 \\\hline
dihedral symmetry $+$ final entry       &  5 &  24 &        140 \\\hline
next-to-final entry                     &  3 &  12 &         57 \\\hline
collinear limit                         &  0 &   1 &         14 \\\hline
$L^{\rm th}$ discontinuity                &  0 &   1 &         13 \\\hline
OPE $T^2\,\ln^{L-1}T$                    &  0 &   0 &         10 \\\hline
OPE $T^2\,\ln^{L-2}T$                    &  0 &   0 &          4 \\\hline
OPE $T^2\,\ln^{L-3}T$                    &  0 &   0 &          0 \\\hline\hline
\end{tabular}
\caption{Number of parameters left when bootstrapping the form factor
  $\EE^{(L)}$ at $L$-loop order in the function space $\cC$ at symbol level.}
\label{tab:Cparameters}
\end{table}

It will also be important to explore the implications of cosmic Galois
theory, or the coaction principle, for the function space $\cC$.
This space shares with the space of hexagon functions
${\cal H}$~\cite{Caron-Huot:2019bsq} the property of not including 
$\zeta_2$ or $\zeta_3$ as independent constant functions.
At weight four, $\zeta_4$ is an independent constant function
in both $\cC$ and ${\cal H}$.  At weight five, neither $\zeta_5$
nor $\zeta_2\zeta_3$ are independent constant functions in ${\cal H}$.
For $\cC$, we do not know yet whether $\zeta_5$ or $\zeta_2\zeta_3$ are
likewise locked to other functions.
Determining this may require either the construction of the six-loop
form factor $\EE^{(6)}$, or else a deeper understanding of the function space $\cC$.

In addition to probing the coaction principle in bulk kinematics, we can
investigate what constants appear at special points in this space.
For example,
in the case of ${\cal H}$, there is a natural base point $(u,v,w)=(1,1,1)$
where all of the weight-three functions vanish, so $\zeta_3$ does not appear.  
This fact leads to strong coaction-principle constraints
on the MZVs that can appear in higher-weight functions in ${\cal H}$ when evaluated
at $(1,1,1)$~\cite{Caron-Huot:2019bsq}.
Unfortunately, we have not yet found a point analogous to $(1,1,1)$
in the $(u,v)$ plane for $\cC$, where we can look for missing zeta values.
There also may be a higher-loop constant normalization factor
needed to maintain the coaction principle,
analogous to the constant $\rho$ in the case of hexagon 
functions~\cite{Caron-Huot:2019bsq}, but this issue is hard to assess
without the analog of the point $(1,1,1)$.


\section{Numerical Results and Simplifications in Kinematic Limits}
\label{sec:results}

Let us now analyze the numerical and analytical behavior of the form factor
remainder function found in the previous section. We first focus on the
$(u,u,1-2u)$ and $(1,v, -v)$ lines, and then consider various points on
these lines where the form factor evaluates to MZVs or alternating sums.

 
\subsection{Values on particular lines}
\label{sec:linevalues}

Perhaps the simplest place to plot the remainder function is on one of the three
equivalent lines bisecting the Euclidean triangle $0 < u,v,w < 1$ in
\Fig{Fig:uvplane}, which run from one corner to the mid-point
of the opposite side.
For example, one of these line segments is given by setting $v=u$.
The remainder function is then forced to vanish as $u\to0$
(a corner) and also as $u \to 1/2$, since $w \to 0$ at that point (the midpoint
of the opposite side).  By symmetry, the maximum of the absolute value of
the $L$-loop remainder function for $0 < u < 1/2$ occurs at
 the dihedrally symmetric point $u=1/3$, since $u=v=w=1/3$ there.
The numerical values at this point, and the ratios to the previous loop
order, are given in Table~\ref{tab:R_symmax}.

\begin{table}[!t]
\begin{center}
\begin{tabular}{|l|c|c|}
\hline\hline
$L$ & $R^{(L)}(\frac{1}{3},\frac{1}{3},\frac{1}{3})$
    & $R^{(L)}(\frac{1}{3},\frac{1}{3},\frac{1}{3})
      /R^{(L-1)}(\frac{1}{3},\frac{1}{3},\frac{1}{3})$ \\
\hline\hline
2  & $-$0.1489664398982726  & -- \\
3  & $-$8.3720527811583320   &  $+$56.20093 \\
4  &  166.8943388313494956    & $-$19.93470 \\
5  & $-$2666.61569156241573  &  $-$15.97787 \\
\hline\hline
\end{tabular}
\caption{\label{tab:R_symmax} The value of the $L$-loop remainder
  function at the symmetric point
  $(u,v,w)=(\frac{1}{3},\frac{1}{3},\frac{1}{3})$ in the Euclidean
  region, as well as the ratio to the previous loop order.}
\end{center}
\end{table}

Most quantities in planar ${\cal N}=4$ sYM theory exhibit sign alternation
from one loop order to the next, at least at high loop orders.
The sign alternation is associated with a finite radius of convergence
of the perturbative series, and a pole on the negative $g^2$ axis that
is typically (e.g.~for the cusp anomalous dimension~\cite{Beisert:2006ez})
at $g^2 = -1/16$.  This behavior has been explored to seven loops in the six-point
amplitude~\cite{Caron-Huot:2019vjl}, and to four loops in the seven-point
amplitude~\cite{Dixon:2020cnr}, mainly in the Euclidean region.
Interestingly, Table~\ref{tab:R_symmax} shows that
for the three-point form factor remainder $R$ there is no sign
alternation from two to three loops in the Euclidean region.
We will see this feature in other regions as well.
From three to four loops the sign alternation starts up (also in other regions).
Indeed, the successive loop-order ratio $R^{(4)}/R^{(3)}=-19.934697\ldots$
is not that far from the asymptotic cusp value of $-16$.
The ratio $R^{(5)}/R^{(4)}$ given in Table~\ref{tab:R_symmax}
is extremely close to this asymptotic cusp value.
(So close that it might be an accident.)

\begin{figure}[t]
\centering
\rotatebox{90}{\hspace{0.2\paperwidth}\clap{$\frac{R^{(L)}(u,u,1-2u)}{R^{(L)}(1/3,1/3,1/3)}$}}
         \includegraphics[height=0.4\paperwidth,trim=4cm 0cm 0cm 0.75cm,clip]{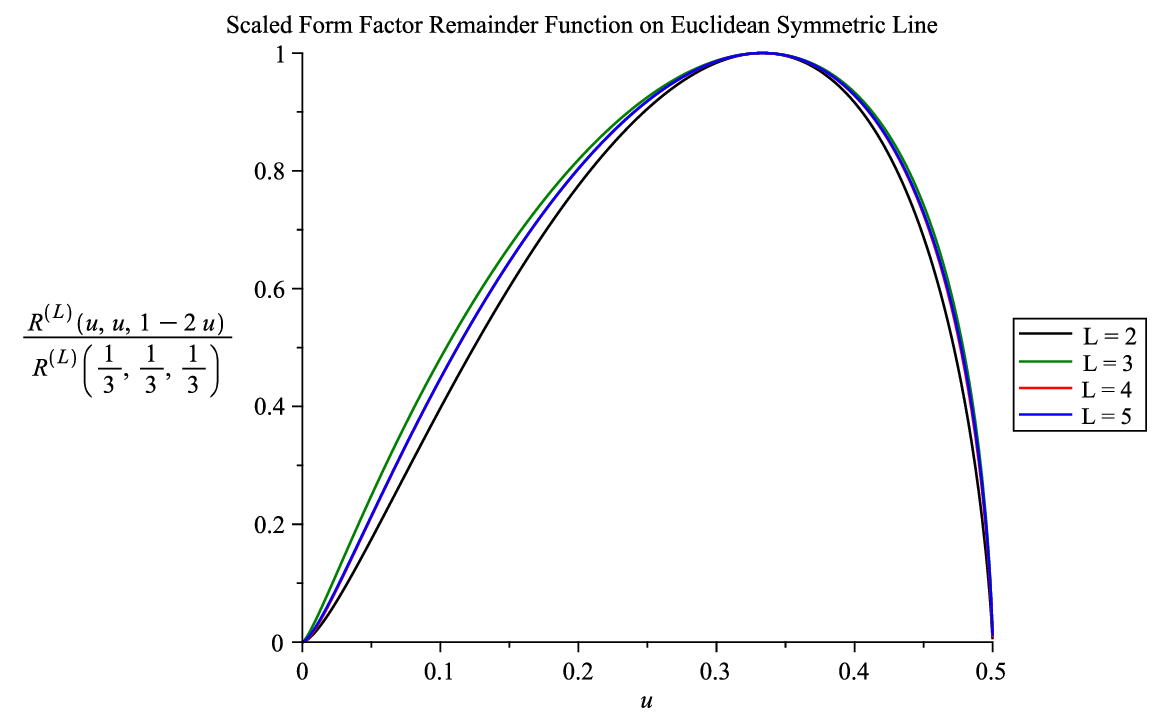}
\caption{The $L$-loop form factor remainder function $R^{(L)}$ on the
  Euclidean line $(u,u,1-2u)$, normalized by its value at $u=1/3$.
  The four-loop curve is hidden under the five-loop one.}
\label{Fig:RLsymline}
\end{figure}

In order to take into account the large dynamical range and lack of
sign alternation, in \Fig{Fig:RLsymline}
we plot the remainder function divided by its maximal value,
given in Table~\ref{tab:R_symmax}.
The $L=4$ curve is hidden directly under the $L=5$ curve, because they are
almost exactly proportional in this region.

We can also study the behavior of the remainder function in scattering
regions, where it becomes complex.
To move from region I to region IIa, we let $\ln w \to \ln|w| - i\pi$
in the vicinity of the line $w=0$.  We then use the coproduct formalism
to integrate up from that line on the IIa branch.
Similarly, to move from region I to region IIIa, we let
$\ln u \to \ln|u| - i\pi$ and $\ln v \to \ln|v| - i\pi$
near the point $u=v=0$, and then integrate up from that point.

\begin{figure}[t]
\centering
 \begin{subfigure}[b]{0.44\textwidth}
         \centering
         \rotatebox{90}{\tiny \hspace{0.12\paperwidth}\clap{$\frac{\text{Re } R^{(L)}(u,u,1-2u)}{\text{Re } R^{(L)}(1,1,-1)}$}}
         \includegraphics[height=0.24\paperwidth,trim=5cm 0cm 2.65cm 0.75cm,clip]{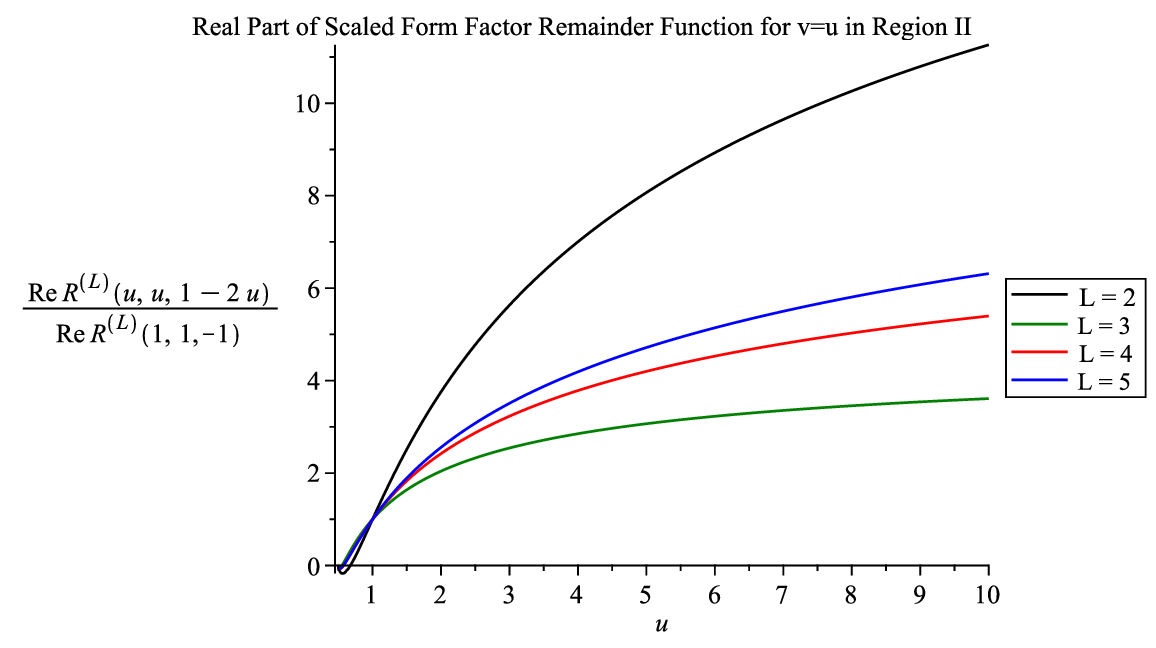}
         \caption{\phantom{.}}
         \label{Fig:ReRLsymII}
     \end{subfigure}
     \hfill
      \begin{subfigure}[b]{0.55\textwidth}
         \centering
         \rotatebox{90}{\tiny \hspace{0.12\paperwidth}\clap{$\frac{\text{Im } R^{(L)}(u,u,1-2u)}{\text{Re } R^{(L)}(1,1,-1)}$}}
         \includegraphics[height=0.24\paperwidth,trim=4.9cm 0cm 0cm 0.75cm,clip]{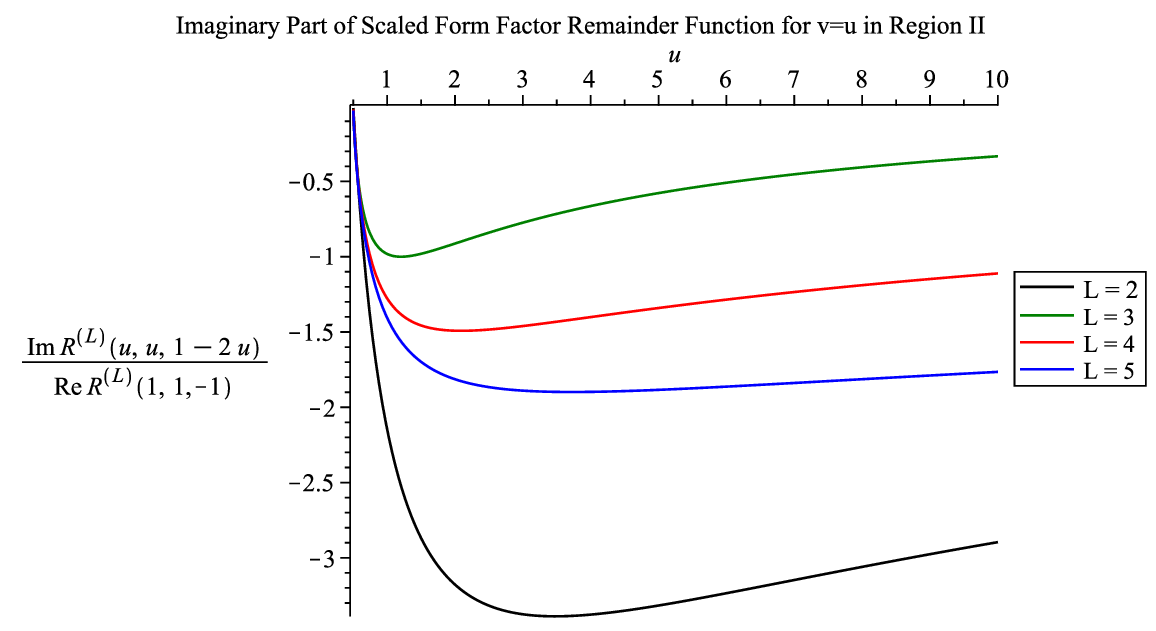}
         \caption{\phantom{.}}
         \label{Fig:ImRLsymII}
     \end{subfigure}
\caption{Real (a) and imaginary (b) part of the
  $L$-loop form factor remainder function $R^{(L)}$ on the line $(u,u,1-2u)$ in scattering region IIa, $u>1/2$,
  normalized by the value of its real part at $u=1$.}
\label{Fig:RLsymII}
\end{figure}

In \Fig{Fig:ReRLsymII},
we plot the real part of the remainder function on the extension of the
line $(u,u,1-2u)$ into the space-like scattering region IIa, for $u>1/2$,
divided by its value at $u=1$. We see that the proportionality of
the four- and five-loop results extends a bit into region IIa, but does not
hold so strongly at larger values of $u$.
In \Fig{Fig:ImRLsymII},
we plot the imaginary part of the remainder function on the same IIa line,
also normalized by the value of its real part at $u=1$.

\begin{figure}[t]
\centering
 \begin{subfigure}[b]{0.475\textwidth}
         \centering
         \rotatebox{90}{\tiny \hspace{0.12\paperwidth}\clap{$\frac{\text{Re } R^{(L)}(-u,-u,1+2u)}{\text{Re } R^{(L)}(-1,-1,3)}$}}
         \includegraphics[height=0.24\paperwidth,trim=5.3cm 0cm 2.5cm 0.675cm,clip]{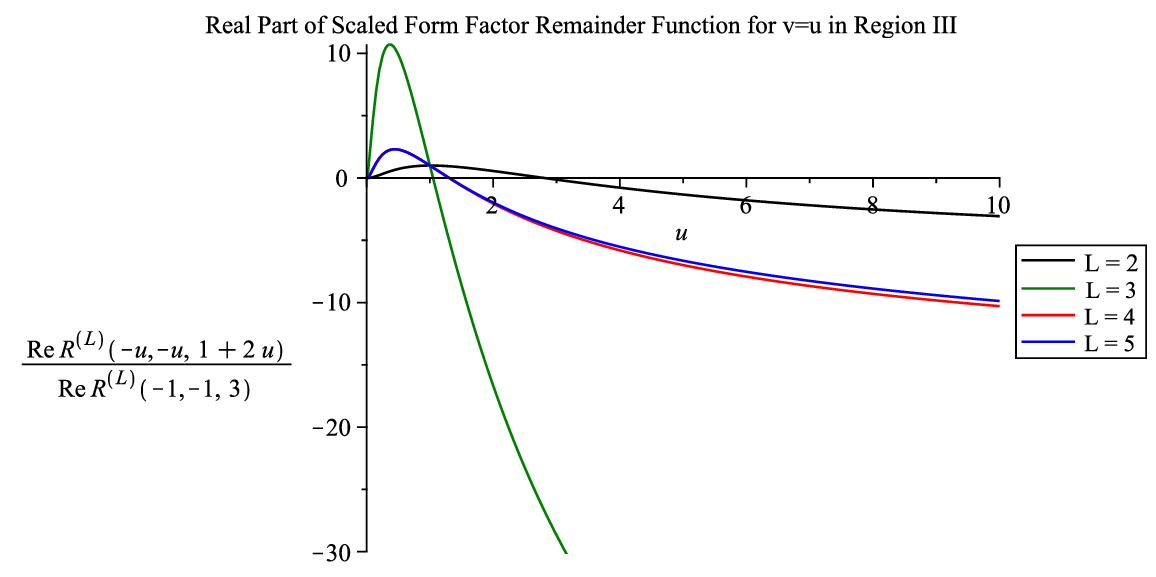}
         \caption{\phantom{.}}
         \label{Fig:ReRLIII}
     \end{subfigure}
     \hfill
      \begin{subfigure}[b]{0.515\textwidth}
         \centering
         \rotatebox{90}{\tiny \hspace{0.12\paperwidth}\clap{$\frac{\text{Im } R^{(L)}(-u,-u,1+2u)}{\text{Re } R^{(L)}(-1,-1,3)}$}}
         \includegraphics[height=0.24\paperwidth,trim=5.5cm 0cm 0cm 0.665cm,clip]{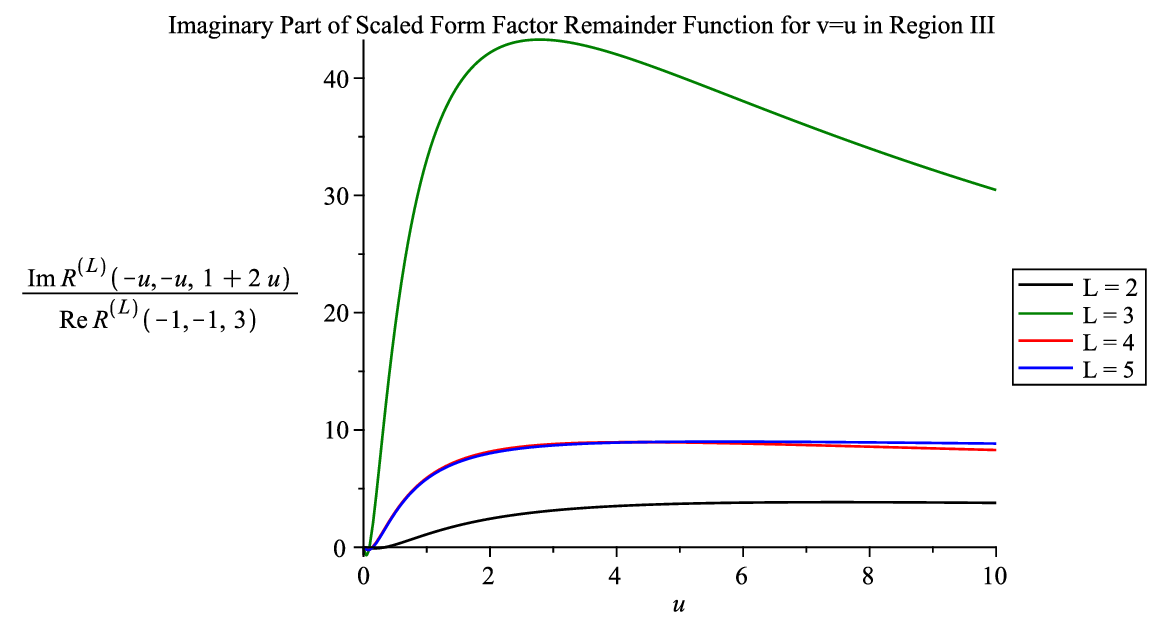}
         \caption{\phantom{.}}
         \label{Fig:ImRLIII}
     \end{subfigure}
\caption{Real (a) and imaginary (b) part of the
  $L$-loop form factor remainder function $R^{(L)}$ on 
  the line $(u,u,1-2u)$ in scattering region IIIa, $u<0$,
  normalized by the value of its real part at $u=-1$.}
\label{Fig:RLIII}
\end{figure}

The analogous plots for the extension of the line $(u,u,1-2u)$ into the
time-like scattering region IIIa, with $u<0$, are given in
\Fig{Fig:ReRLIII} and \Fig{Fig:ImRLIII}. In this case, we normalize
by the value of the real part at $u=-1$.  In this region, the proportionality
of the four- and five-loop remainder function holds to much larger values of $|u|$
than in region II.

\begin{figure}[t]
\centering
 \begin{subfigure}[b]{0.445\textwidth}
         \centering
         \rotatebox{90}{\tiny \hspace{0.12\paperwidth}\clap{$\frac{\text{Re } R^{(L)}(1,v,-v)}{ R^{(L)}(1,\infty,-\infty)}$}}
         \includegraphics[height=0.24\paperwidth,trim=4.2cm 0cm 2.75cm 0.675cm,clip]{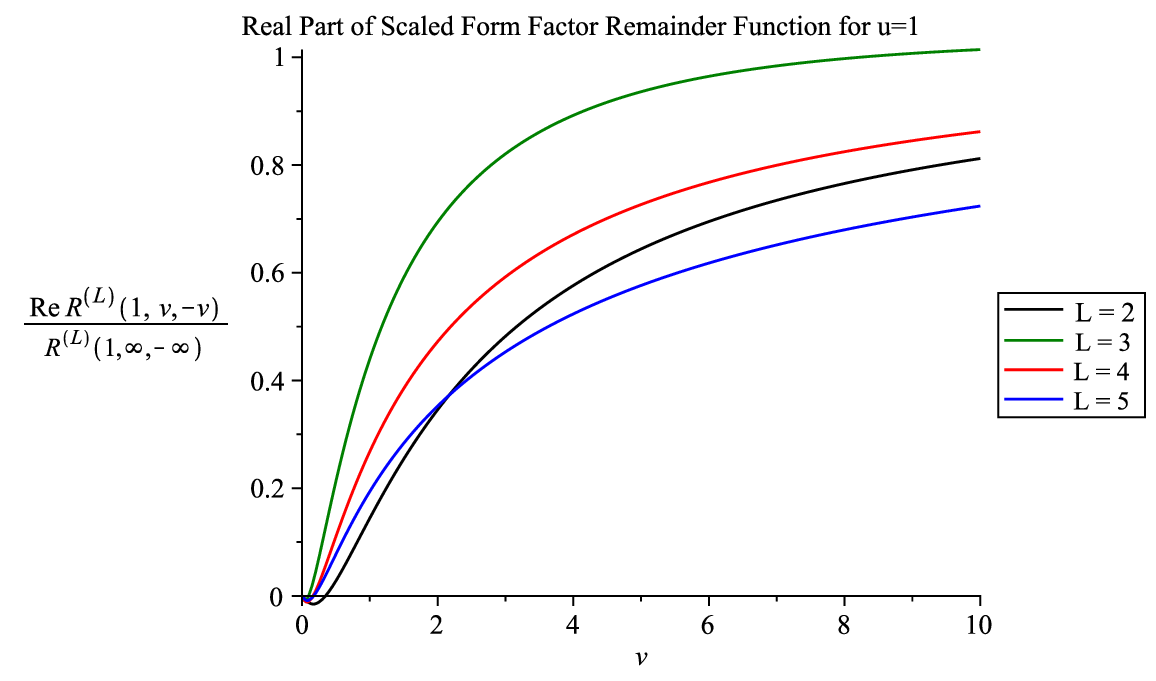}
         \caption{\phantom{.}}
         \label{Fig:ReRLII}
     \end{subfigure}
     \hfill
      \begin{subfigure}[b]{0.54\textwidth}
         \centering
         \rotatebox{90}{\tiny \hspace{0.12\paperwidth}\clap{$\frac{\text{Im } R^{(L)}(1,v,-v)}{ R^{(L)}(1,\infty,-\infty)}$}}
         \includegraphics[height=0.24\paperwidth,trim=4.2cm 0cm 0cm 0.75cm,clip]{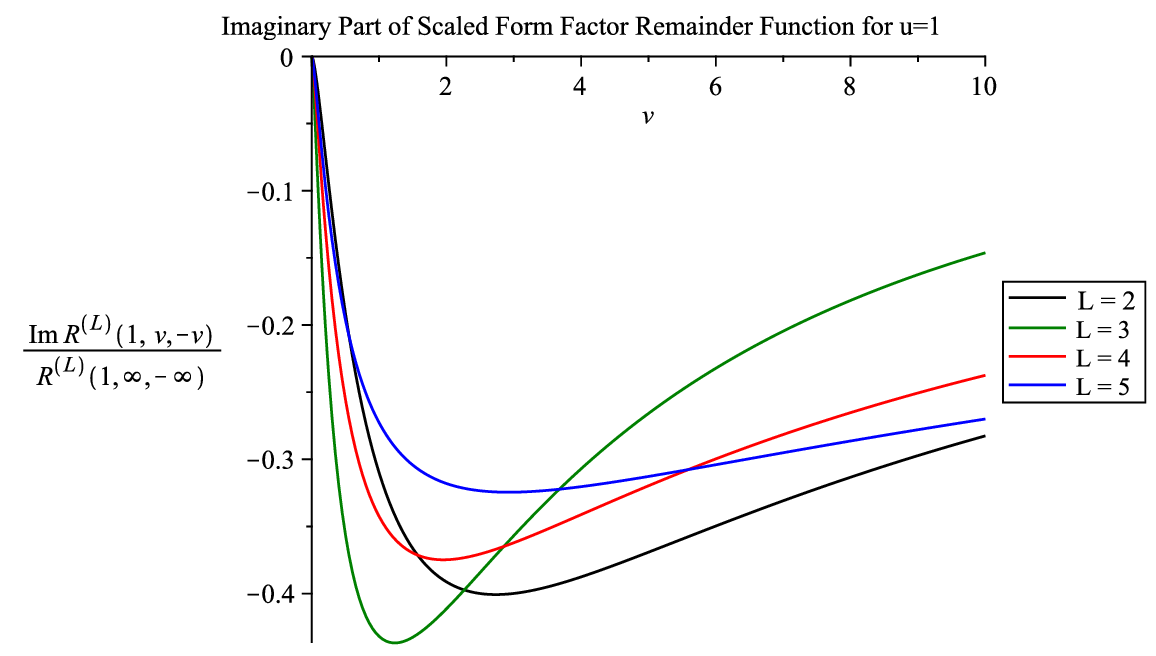}
         \caption{\phantom{.}}
         \label{Fig:ImRLII}
     \end{subfigure}
\caption{Real (a) and imaginary (b) part of the
  $L$-loop form factor remainder function $R^{(L)}$ on the line $u=1$ in scattering region IIa, $v>0$,
  normalized by its (real) value at $v=\infty$.}
\label{Fig:RLII}
\end{figure}

Finally, we plot the remainder function in scattering region IIa again,
but now on the line $u=1$, in \Fig{Fig:ReRLII} and \Fig{Fig:ImRLII}.
In this case the value at $v=\infty$ is real, and we normalize by that value.

\subsection{Values at particular points}
\label{sec:pointvalues}

There are some distinguished points in the $(u,v)$ plane where
the form factor remainder function takes on interesting analytic values.
Of course, $R$ vanishes on the entire boundary of the Euclidean triangle,
which includes the points $(u,v,w) = (1,0,0)$, $(0,1,0)$ and $(0,0,1)$.
For the function space $\cM$ as a whole, we expect only MZVs and
logarithms of the two small variables at these points.

The point $(u,v,w) = (1,1,-1)$ lies at the intersection of the $u=1$
line and the $u=v$ line plotted earlier.  From the representation
of $\cM$ on the $u=1$ line as HPLs of argument $v$ with weight vector
elements $\{0,1,-1\}$, we know that the real and imaginary parts of $R$
at the point $(1,1,-1)$ must belong to the space of alternating sums.
For example, the values at two and three loops are
\bea
R^{(2)}(1,1,-1) &=& 8 \, {\rm Li}_4(1/2) + \frac{1}{3} \, \ln^4 2
+ 4 \, \zeta_2 \, \ln^2 2 - \frac{29}{4} \, \zeta_4
\nonumber\\&&\hskip0.0cm\null
+ i \pi \biggl[ \frac{4}{3} \ln^3 2
          + 2 \, \zeta_2 \, \ln 2 - 2 \, \zeta_3 \biggr] \,,
\label{R2_11}
\eea
\bea
R^{(3)}(1,1,-1) &=& - 96 \, {\rm Li}_6(1/2) + 8 \, \zeta_2 \, {\rm Li}_4(1/2)
- \frac{2}{15} \, \ln^6 2 - \frac{23}{3} \, \zeta_2 \, \ln^4 2
- 104 \, \zeta_4 \, \ln^2 2
\nonumber\\&&\hskip0.0cm\null
+ 24 \, \zeta_2 \, \zeta_3 \, \ln 2 + \frac{76}{3} \,\zeta_6 + 8 \, (\zeta_3)^2
\nonumber\\&&\hskip0.0cm\null
+ i \, \pi \, \biggl[ 32 \, {\rm Li}_5(1/2) - \frac{2}{5} \, \ln^5 2
  - \frac{4}{3} \, \zeta_2 \, \ln^3 2 - 4 \, \zeta_3 \, \ln^2 2
\nonumber\\&&\hskip0.7cm\null
  +  \Bigl( 16 \, {\rm Li}_4(1/2) - 19 \, \zeta_4 \Bigr) \, \ln 2
          + \frac{1}{4} \, \zeta_5 + 2 \, \zeta_2 \, \zeta_3 \biggr] \,. 
\label{R3_11}
\eea
The values at four and five loops can be written in terms of an
alternating-sum $f$-basis using
{\sc HyperlogProcedures}~\cite{HyperlogProcedures},
but the results are fairly lengthy.
In Table~\ref{tab:R_11} we present the numerical values of $R^{(L)}$
and of the ratio of successive loop orders $R^{(L)}/R^{(L-1)}$ at this point.

\begin{table}[!t]
\begin{center}
\begin{tabular}{|l|c|c|}
\hline\hline
$L$ & $R^{(L)}(1,1,-1)$ & $R^{(L)}(1,1,-1)/R^{(L-1)}(1,1,-1)$ \\
\hline\hline
2  &  $-$0.4688118 + 1.0061952 $i$ & -- \\
3  & $-$28.364132 + 27.844487 $i$ & $+$33.528634 + 12.567654 $i$ \\
4  &  546.09247 $-$ 697.38435 $i$ & $-$22.095756 + 2.8958875 $i$ \\
5  & $-$8052.1673 + 11322.282 $i$ & $-$15.668885 + 0.7234070 $i$ \\
\hline\hline
\end{tabular}
\caption{\label{tab:R_11} The value of the remainder
  function at $(u,v,w)=(1,1,-1)$ at loop order $L$, as well as
  the ratio to the previous loop order.  The five loop to four loop
  ratio is already not far from the asymptotic ratio of $-16$
  for the cusp anomalous dimension.}
\end{center}
\end{table}

When we take the limit $v\to\infty$ on the $u=1$ line,
the remainder function becomes real, and the values belong
to the space of MZVs:
\bea
R^{(2)}(1,\infty,-\infty) &=& - 3 \zeta_4 \,,
\label{R2_vtoinf}\\
R^{(3)}(1,\infty,-\infty) &=&
- \frac{931}{12} \, \zeta_6 + 10 \, (\zeta_3)^2 \,,
\label{R3_vtoinf}\\
R^{(4)}(1,\infty,-\infty) &=& \frac{22193}{9} \, \zeta_8
- 352  \,  \zeta_3  \,  \zeta_5 - 20  \, \zeta_{5,3} \,,
\label{R4_vtoinf}\\
R^{(5)}(1,\infty,-\infty) &=&
- \frac{941509787}{16800} \, \zeta_{10} - 493 \, \zeta_4 \, (\zeta_3)^2
+ 920 \, \zeta_2 \, \zeta_3 \, \zeta_5
+ \frac{29049}{7} \, (\zeta_5)^2
\nonumber\\&&\hskip0.0cm\null
+ \frac{14843}{2} \, \zeta_3 \, \zeta_7
+ \frac{124}{5} \, \zeta_2 \, \zeta_{5,3} + \frac{1416}{7} \, \zeta_{7,3} \,.
\label{R5_vtoinf}
\eea
We provide the numerical values, and the successive loop order ratios,
in Table~\ref{tab:R_1inf}.  The approach to the asymptotic cusp ratio
of $-16$ is slower out at infinity than it is closer to the origin,
which is similar to what is observed for six-point
amplitudes~\cite{Caron-Huot:2019vjl}.

\begin{table}[!t]
\begin{center}
\begin{tabular}{|l|c|c|}
\hline\hline
$L$ & $R^{(L)}(1,\infty,-\infty)$
    & $R^{(L)}(1,\infty,-\infty)/R^{(L-1)}(1,\infty,-\infty)$ \\
\hline\hline
2  & $-$3.2469697   & -- \\
3  & $-$64.479458   & $+$19.858349 \\
4  & 2036.439992    & $-$31.582771 \\
5  & $-$41521.8935  & $-$20.389451 \\
\hline\hline
\end{tabular}
\caption{\label{tab:R_1inf} The value of the remainder
  function at $(u,v,w)=(1,\infty,-\infty)$, i.e.~the $v\to +\infty$
  limit of the $u=1$ line, at loop order $L$, as well as
  the ratio to the previous loop order.  The five-loop to four-loop ratio
  is not quite as close to the cusp asymptotic value of $-16$
  as was the case at $(u,v,w)=(1,1,-1)$.}
\end{center}
\end{table}

A second limit in which the remainder function becomes real and
evaluates to MZVs is the limit where $u\to\pm\infty$ on the line $(u,u,1-2u)$.
The value is independent of the sign of $u$; that is, it is the same whether we approach 
this point from within region IIa or region IIIa.  Through five loops, the limit is
\bea
R^{(2)}(u,u,1-2u)\Big|_{u\to\infty} &=& - 10 \, \zeta_4 \,,
\label{R2_symutoinf}\\
R^{(3)}(u,u,1-2u)\Big|_{u\to\infty} &=&
- \frac{371}{3} \, \zeta_6 + 8 \, (\zeta_3)^2 \,,
\label{R3_symutoinf}\\
R^{(4)}(u,u,1-2u)\Big|_{u\to\infty} &=&
  5212 \, \zeta_8 + 64 \, \zeta_2 \, (\zeta_3)^2
- 160 \, \zeta_3 \, \zeta_5 - 24 \, \zeta_{5,3} \,,
\label{R4_symutoinf}\\
R^{(5)}(u,u,1-2u)\Big|_{u\to\infty} &=&
- \frac{600943}{5} \, \zeta_{10} - 672 \, \zeta_4 \, (\zeta_3)^2
- 1280 \, \zeta_2 \, \zeta_3 \, \zeta_5
+ 708 \, (\zeta_5)^2
\nonumber\\&&\hskip0.0cm\null
+ 1680 \, \zeta_3 \, \zeta_7
+ 240 \, \zeta_2 \, \zeta_{5,3} + 192 \, \zeta_{7,3} \,.
\label{R5_symutoinf}
\eea
The numerical values and the ratios to the previous loop order are
shown in Table~\ref{tab:R_syminf}.  The approach to the asymptotic
cusp ratio is even slower here.

\begin{table}[!t]
\begin{center}
\begin{tabular}{|l|c|c|}
\hline\hline
$L$ & $R^{(L)}(u,u,1-2u)\Big|_{u\to\infty} $
    & $R^{(L)}(u,u,1-2u)/R^{(L-1)}(u,u,1-2u)\Big|_{u\to\infty}$ \\[.12cm]
\hline\hline
2  & $-$10.823232   & -- \\
3  & $-$114.25190   &  $+$10.556172 \\
4  & 5185.0320798    & $-$45.382458 \\
5  & $-$121169.4098  & $-$23.369076 \\
\hline\hline
\end{tabular}
\caption{\label{tab:R_syminf} The value of the $L$-loop remainder
  function at $(u,v,w)=(u,u,1-2u)$ with $u\to\infty$,
  as well as the ratio to the previous loop order.}
\end{center}
\end{table}

The general conclusion of our numerical investigation is a rough consistency
with the expected large-order asymptotics.  The two-loop remainder function
is a bit smaller, and has the opposite sign, than one might have expected
from the higher-loop results, and the more strictly
sign-alternating behavior of six-and seven-point amplitude remainder
functions~\cite{Dixon:2013eka,Dixon:2014voa,Caron-Huot:2019vjl,Dixon:2020cnr}.
Also, the form factor remainder function tends to a constant at infinity,
which has interesting consequences for the high-energy or Regge limit,
as we will mention in section~\ref{sec:conclusions}.

\section{Multi-Loop One-Mass Four-Point Integrals}
\label{sec:generalintegrals}

As mentioned in section~\ref{adjrestrict}, the ES-like adjacency
restriction~(\ref{eq:ExtSteinmannomuomv}), which is critical to defining the
function space $\cM$, was first motivated by its appearance in
the form factor (as opposed to the remainder function) computed in
ref.~\cite{Brandhuber:2012vm}.
We have also inspected several other three-point form factors,
or amplitudes involving a Higgs boson and three gluons,
that are available in the
literature~\cite{Gehrmann:2011aa,Brandhuber:2012vm,%
  Brandhuber:2014ica,Brandhuber:2017bkg}. All obey the same ES-like restriction.
The constraint~(\ref{eq:ExtSteinmannomuomv}) has
no effect until weight three, because $1-u$ cannot appear until the
second entry, but it can be seen in the weight-three and -four functions
in these references.

All of the quantities in question are linear combinations
of a class of two-loop
master integrals associated with four-point processes
with one massive and three massless external legs, and all massless
internal lines, for both planar and non-planar topologies.
The master integrals for these topologies were first computed
through weight four in refs.~\cite{Gehrmann:2000zt,Gehrmann:2001ck}.
More recently they have been expanded further in the dimensional
regularization parameter $\epsilon$, through weights
five~\cite{Duhr:2014nda}
and six~\cite{Mistlbergerprivate}. Examining
these two-loop master integrals through weight six~\cite{Mistlbergerprivate},
we find that they all lie within $\cM$.\footnote{This was checked at function level through weight five, but only at symbol level at weight six.}
There are 89 different master integrals.  At weight four,
69 symbols and 79 functions of the 89 are linearly independent,
while there are $3^4 = 81$ symbols and (from Table~\ref{tab:FvsC}) 94
functions in $\cM_4$.
Thus only 12 of the weight-four symbols in $\cM$ are not captured by the
two-loop master integrals.
At weight five, there are 72 and 86 independent symbols and functions,
respectively. (These numbers depend a bit on how the integrals
are normalized; shifting them all by factors like $(u_i)^{\e}$ can
change the number of independent higher-weight functions.)
We expect much more of the space $\cM$ at weights five and six
to be filled out by three-loop integrals.

Based on the fact that all the two-loop master integrals are in $\cM$, 
as well as the planar ${\cal N}=4$ sYM form factors through five loops, 
we expect that this space contains all master integrals for the same 
topologies, one massive and three massless external legs, and all 
massless external lines, planar and non-planar, to higher loop orders.  
However, for reasons discussed in the ``note added'', we don't expect 
this statement to be true to arbitrary loop order. If our expectation is 
true, then any four-point amplitude with these kinematics, for example 
subleading-in-$N$ corrections in ${\cal N}=4$ sYM, as well as QCD 
corrections to amplitudes for $gg\to Hg$ (in the heavy-top limit), 
$qg\to Zq$, and $e^+e^- \to q\bar{q}g$, can be expressed to three loops, possibly further,
in terms of the transcendental functions in $\cM$.  All different weights 
up to $2L$ can be expected to appear at $L$ loops in QCD,
and the rational-function prefactors in front of these functions, especially 
the lower-weight ones, can be expected to become rather intricate. 
Nevertheless, this would be a remarkable property for a large class of 
high-order amplitudes in massless theories to exhibit.

\section{Conclusions and outlook}
\label{sec:conclusions}

In this paper, we have bootstrapped the three-point form factor of the chiral part of the stress-tensor supermultiplet in planar $\mathcal{N}=4$ sYM theory through five loops, extending the state of the art by three loop orders. To carry out this bootstrap, we have utilized new boundary data coming from the OPE in the near-collinear limit of these form factors~\cite{Sever:2020jjx,Toappear1,Toappear2}. The OPE also provides important cross-checks on our results.  On the other hand, the information about subleading logarithms in $T$ at order $T^2$ coming from our results also assisted the determination of the higher-order behavior of the form factor transition.  The integrability-based bootstrap and the perturbative form factor bootstrap thus complement each other quite nicely.

These new form factor results provide a novel glimpse into the mathematical structure of gauge theory at high loop orders, complementing our rapidly-developing understanding of the mathematical structure of multi-loop amplitudes. In particular, we have observed novel extended-Steinmann-like constraints~\eqref{eq:ExtSteinmannomuomv} that are obeyed by the finite part of the form factor $\mathcal{E}^{(L)}$. Similar to the extended Steinmann relations for amplitudes~\cite{Caron-Huot:2019bsq}, we do not have a full proof of these relations based on physical principles. It would thus be interesting to elucidate the origin of these restrictions, perhaps using the types of methods employed in refs.~\cite{Abreu:2014cla,Bloch:2015efx,Abreu:2017ptx,Bourjaily:2019exo,Bourjaily:2020wvq,Benincasa:2020aoj}.

We have also observed that the functions $\mathcal{E}^{(L)}$ satisfy multiple-final-entry conditions. Namely, the double coproduct entries of $\mathcal{E}^{(L)}$ satisfy relation~\eqref{maindoublefinalrelation}, and its dihedral images, through at least five loops. It would be interesting to see if these relations could be derived using the $\bar{Q}$ equation~\cite{CaronHuot:2011ky,CaronHuot:2011kk}. Similarly, the triple coproduct entries of $\mathcal{E}^{(L)}$ satisfy relations~\eqref{FE3eq1}--\eqref{FE3eq5}, and their dihedral images, through the same loop order. It is worth noting that the six-point amplitude also satisfies next-to-final-entry conditions~\cite{Dixon:2015iva}; however, when the extended Steinmann relations are used to build up the space of hexagon functions, these conditions are automatically satisfied~\cite{Caron-Huot:2019vjl}. As seen in Table~\ref{tab:Mparameters} and Table~\ref{tab:Cparameters}, the multiple-final-entry constraints that we observe are not redundant with the other general assumptions we currently build into our ansatz. 

In addition to exhibiting interesting mathematical structure, these form factors should be closely related to quantities of phenomenological interest. In particular, we expect the remainder function $R^{(L)}$ to match the maximally transcendental parts of the $gg\to Hg$ and $H\to ggg$ amplitude remainder functions in the heavy-top limit of QCD. Additionally, for the reasons described in section~\ref{sec:generalintegrals}, we believe that the lower-transcendentality functions appearing in these QCD amplitudes at three loops will be contained in the form factor space $\cM$, although they will have more complicated rational prefactors. These form factors, and the function space $\cM$, are thus important for future studies of the Higgs boson, among other LHC processes. 

An interesting avenue for future work is to study the high-energy
or Regge behavior of four-point scattering involving a colorless operator.
In the case of four-gluon scattering, where the remainder function vanishes,
the BDS formula~\cite{Bern:2005iz}
is consistent with Regge behavior being controlled
by the cusp and collinear anomalous
dimensions~\cite{Drummond:2007aua,Naculich:2007ub}.
In the present case, the remainder function does not vanish in the
scattering region II (space-like operator), but
as we saw in section~\ref{sec:pointvalues}, it goes to a real constant as
$v \to \infty$ for $u=1$, or for $u=v$.  This constancy follows from the
final-entry condition, $R^{u_i} + R^{1-u_i} = 0$.
For $u\ll v$, this region
is the high-energy limit $s_{23} \gg s_{12}, q^2$.  The logarithmic growth
with energy again is captured by the infrared-divergent terms,
as well as $E^{(1)}$, which have been removed from the remainder function.
However, there will also be an
impact factor, which does not depend on $s_{23}$, but does depend
on the dimensionless ratio $u = s_{12}/q^2$. It can be extracted from the remainder function to high loop orders.
Similar remarks apply to scattering region III (the time-like operator).

We have defined the form factor function space $\cM$ to have the analytic properties expected of a generic Feynman integral that draws from the $\mathcal{S}_3$ symbol alphabet, and to obey the extended-Steinmann-like relations~\eqref{eq:ExtSteinmannomuomv}. In order to carry out our bootstrap, we have iteratively constructed $\cM$ through weight eight. While the space $\cM$ involve fewer symbol letters than the hexagon function space $\mathcal{H}$, we find that its dimension grows faster with the transcendental weight $w$, namely as $3^w$ --- compared to $\sim 1.8^w$ in the hexagon case. However, the clean growth of the function space with the weight and the simpler letters make $\cM$ an ideal testing ground for constructing the function space in a closed form.

In fact, we have found that the function $\mathcal{E}$ is contained in the smaller subspace $\cC \subset \cM$. We expect this space to be crucial for going beyond five-loop order \cite{in-progress}, but we do not yet have a bottom-up, or first-principles, construction of $\cC$.

It would also be interesting to bootstrap form factors with a larger number of points $n$. In particular, a non-trivial NMHV form factor is first encountered at $n=4$.  Higher-point form factors for $\tr(F_{SD}^2)$ factorize onto $n$ gluon scattering amplitudes for $n\geq4$.  Since these amplitudes do not obey a maximal-transcendentality principle, we also do not expect such a principle to hold between ${\cal N}=4$ sYM form factors and QCD Higgs amplitudes, beyond $n=3$.

Finally, it would be interesting to bootstrap form factors of different operators. 
Possible operators include those studied in refs.~\cite{Bork:2010wf,Engelund:2012re,Brandhuber:2014ica,Wilhelm:2014qua, Nandan:2014oga, Loebbert:2015ova, Brandhuber:2016fni, Loebbert:2016xkw, Caron-Huot:2016cwu,Ahmed:2016vgl,Banerjee:2016kri,Huber:2019fea,Lin:2020dyj}. In particular, in the case of the two-loop three-point form factor for $\tr(F^3)$ it has been shown that the next-to-leading order contribution to the Higgs amplitude remainder in the heavy-top limit of QCD shares its maximally transcendental part with its counterpart in $\mathcal{N}=4$ sYM theory~\cite{Brandhuber:2017bkg,Brandhuber:2018xzk,Brandhuber:2018kqb,Jin:2018fak}. These form factors are liable to have just as interesting a structure at higher loops as the form factor we have studied in this work.

\begin{quote}
{\bf Note Added:} The initial version of this paper included the conjecture that the transcendental functions appearing in four-point one-mass amplitudes would be contained in the $\mathcal{M}$ space to all loop orders. After it appeared, we were reminded by Erik Panzer that non-polylogarithmic periods are known to appear in massless $\phi^4$ theory by eight loops~\cite{Brown:2010bw,Brown:2013wn,Panzer:2016snt}.  Because these period integrals can be converted to propagator integrals by injecting external momentum, and since propagator integrals are contained within four-point one-mass integrals, the polylogarithmic conjecture we originally put forward must fail at higher loop orders (in generic massless quantum field theories).
\end{quote}

\acknowledgments
We would like to thank Amit Sever, Alexander Tumanov, Benjamin Basso,
Claude Duhr, Chi Zhang, and Erik Panzer for stimulating discussions.
We especially thank Bernhard Mistlberger for providing the two-loop
master integrals through weight six.
This research was supported by the US Department of Energy under contract
DE--AC02--76SF00515.
AJM and MW were supported in part by the ERC starting grant 757978 and grant 00015369 from the Villum Fonden. AJM was additionally supported by a Carlsberg Postdoctoral Fellowship
(CF18-0641), and MW was additionally supported by grant 00025445 from Villum Fonden.

\appendix

\section{Pentaladder integrals}  \label{app:pentaladders}

The pentaladder integrals, or pentabox ladder integrals, belong to both the
form factor space $\cM$ and the heptagon function space. As such, they provide
a way of understanding the ES-like adjacency restrictions for some of the functions
in $\cM$, in terms of adjacency restrictions which follow from extended
Steinmann relations for heptagon functions, as we will discuss in this appendix.

\begin{figure}[t]
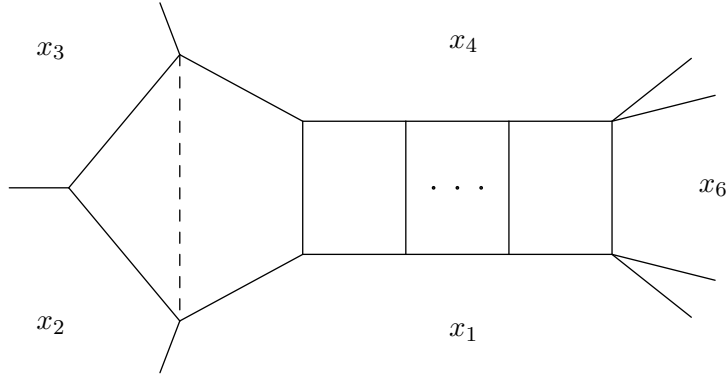

\begin{center}
\begin{fmfchar*}(300,140)
	\fmfset{dot_size}{0.18mm}
	 %
	\fmfforce{(.115w,.5h)}{p1}
	\fmfforce{(.255w,.86h)}{p2}
	\fmfforce{(.41w,.68h)}{p3}
	\fmfforce{(.54w,.68h)}{p4}
	\fmfforce{(.67w,.68h)}{p5}
	\fmfforce{(.8w,.68h)}{p7}
	\fmfforce{(.8w,.32h)}{p9}
	\fmfforce{(.67w,.32h)}{p11}
	\fmfforce{(.54w,.32h)}{p12}
	\fmfforce{(.41w,.32h)}{p13}
	\fmfforce{(.255w,.14h)}{p14}
	%
	\fmfforce{(.575w,.5h)}{d1}
	\fmfforce{(.605w,.5h)}{d2}
	\fmfforce{(.635w,.5h)}{d3}
	%
	\fmfforce{(.04w,.5h)}{e1}
	\fmfforce{(.23w,1.0h)}{e2}
	\fmfforce{(0.9w,0.85h)}{e3}
	\fmfforce{(0.93w,0.75h)}{e4}
	\fmfforce{(0.93w,0.25h)}{e5}
	\fmfforce{(0.9w,0.15h)}{e6}
	\fmfforce{(.23w,0.0h)}{e7}
	%
	\fmfforce{(.12w,.84h)}{l1}
	\fmfforce{(.59w,.85h)}{l2}
	\fmfforce{(0.9w,.5h)}{l34}
	\fmfforce{(.59w,.15h)}{l5}
	\fmfforce{(.12w,.16h)}{l6}
	%
	\fmf{plain, width=.2mm}{p1,p2}
	\fmf{plain, width=.2mm}{p2,p3}
	\fmf{plain, width=.2mm}{p3,p4}
	\fmf{plain, width=.2mm}{p4,p5}
	\fmf{plain, width=.2mm}{p5,p7}
	\fmf{plain, width=.2mm}{p7,p9}
	\fmf{plain, width=.2mm}{p9,p11}
	\fmf{plain, width=.2mm}{p11,p12}
	\fmf{plain, width=.2mm}{p12,p13}
	\fmf{plain, width=.2mm}{p13,p14}
	\fmf{plain, width=.2mm}{p14,p1}
	%
	\fmf{dashes, width=.2mm}{p2,p14}
	%
	\fmf{plain, width=.2mm}{p3,p13}
	\fmf{plain, width=.2mm}{p4,p12}
	\fmf{plain, width=.2mm}{p5,p11}
	%
	\fmf{plain, width=.2mm}{p1,e1}
	\fmf{plain, width=.2mm}{p2,e2}
	\fmf{plain, width=.2mm}{p7,e3}
	\fmf{plain, width=.2mm}{p7,e4}
	\fmf{plain, width=.2mm}{p9,e5}
	\fmf{plain, width=.2mm}{p9,e6}
	\fmf{plain, width=.2mm}{p14,e7}
	%
	\fmfdot{d1}
	\fmfdot{d2}
	\fmfdot{d3}
	%
	\fmfv{label={$x_3$}, label.dist=.1cm}{l1}
	\fmfv{label={$x_4$}, label.dist=.1cm}{l2}
	\fmfv{label={$x_6$}, label.dist=.1cm}{l34}
	\fmfv{label={$x_1$}, label.dist=.1cm}{l5}
	\fmfv{label={$x_2$}, label.dist=.1cm}{l6}
\end{fmfchar*}
\end{center}
\caption{The seven-point pentabox ladder integral, labeled by dual coordinates, in which the box ladder involves $L-1$ loops. The dashed line represents a numerator factor that renders the integral DCI and infrared finite. This class of integrals maps into the space of functions relevant for the three-point form factor in the limit $x_6 \to \infty$.}
\label{fig:pentaladder}
\end{figure}

The class of pentaladder integrals depicted in \Fig{fig:pentaladder} was first considered in ref.~\cite{Drummond:2010cz}. With the inclusion of a numerator associated with the pentagon loop (which is depicted graphically by a dashed line), they are rendered infrared finite and dual conformally invariant (DCI).
Starting from the one-loop pentagon integral,
\be
{\cal I}_{pl}^{(1)}(x_1,x_2,x_3,x_4,x_6) = \frac{x_{14}^2 x_{26}^2 x_{36}^2 }{x_{6a}^2} \int \frac{d^4 x_r}{i \pi^2} \frac{x_{ar}^2}{x_{1r}^2 x_{2r}^2 x_{3r}^2 x_{4r}^2 x_{6r}^2} \, ,
\label{eq:pent1}
\ee
where $x_a$ is the dual point that satisfies the null-separation conditions $x_{a1}^2 = x_{a2}^2 = x_{a3}^2 = x_{a4}^2 = 0$, the $L$-loop pentaladder integral can be constructed iteratively by including $L{-}1$ further integrations of the form
\be
{\cal I}_{pl}^{(L)}(x_1,x_2,x_3,x_4,x_6) = \frac{x_{14}^2 x_{26}^2 x_{36}^2 }{x_{6a}^2}\int \frac{d^4 x_r}{i \pi^2} \frac{x_{ar}^2 }{x_{1r}^2 x_{2r}^2 x_{3r}^2 x_{4r}^2 x_{6r}^2} {\cal I}_{pl}^{(L-1)}(x_1,x_2,x_3,x_4,x_r) \, .
\ee
These integrals depend on the pair of cross-ratios
\be \label{eq:pentaladder_cross_ratios}
U = \frac{x_{46}^2 x_{13}^2}{x_{36}^2 x_{14}^2} \,, \qquad
V = \frac{x_{16}^2 x_{24}^2}{x_{26}^2 x_{14}^2}  \,, 
\ee
which describe a two-dimensional subspace of seven-particle kinematics. 


An important aspect of this class of integrals is that adjacent loop orders are related to each other by a second-order differential equation~\cite{Drummond:2010cz}. Specifically, defining the function
\be\label{eq:PsiDef}
\Psi^{(L)}(U,V) = (1-U-V) \, {\cal I}_{pl}^{(L)}(x_1,x_2,x_3,x_4,x_6)\,,
\ee
we have that
\be
U V (1-U-V)\del_V\del_U \Psi^{(L)}(U,V)= \Psi^{(L-1)}(U,V) \, .
\label{PsiDiffeq}
\ee
Using this relation, these integrals have been computed to high loop order and resummed in the coupling~\cite{Caron-Huot:2018dsv}. As a result, explicit polylogarithmic expressions for these integrals are available through eight loops. 

These integrals also turn out to be relevant to the space of functions entering the three-point form factors in this paper. To see how, consider the limit in which the dual point $x_6$ is sent to infinity. In this limit, the seven-point cross ratios~\eqref{eq:pentaladder_cross_ratios} simplify to ratios that only depend on three external momenta:
\begin{align} \label{eq:x6_inf_lim}
U \xrightarrow[]{x_6 \to \infty}  \frac{x_{13}^2}{x_{14}^2} = \frac{s_{12}}{s_{123}} = u \, , \qquad
V \xrightarrow[]{x_6 \to \infty}  \frac{x_{24}^2}{x_{14}^2} = \frac{s_{23}}{s_{123}} = v \, . 
\end{align}
Thus, these variables map directly to the variables $u$ and $v$ defined in \eqn{uvwdef}, while all kinematic dependence on the box end of the ladders drops out.

Since the variables $U$ and $V$ each remain finite in the limit~\eqref{eq:x6_inf_lim}, the functional form of $\Psi^{(L)}$ does not change, and we can directly check whether they have the right properties to contribute to the form factor $\mathcal{F}_{3}^{\rm MHV}$. As it turns out, they only draw from the five-letter subset 
\be \label{eq:pentaladder_alphabet}
 \{ u,\, v, \, 1-u, \, 1-v, \, 1-u-v \}
 \ee
 of the six-letter symbol alphabet~\eqref{alphabet2}. Also, they obey the same extended Steinmann constraints as the form factor space $\cM$, insofar as the two letters $1{-}u$ and $1{-}v$ never appear sequentially in their symbols. Since only the letters $u$ and $v$ can appear in the first entry in seven-particle kinematics, these functions also satisfy the first-entry condition relevant for form factors. Thus, we see that the functions $\Psi^{(L)}(u,v)$ have all of the right properties to appear in the space of functions $\cM$ constructed in section~\ref{sec:func_space}.

The differential equation~(\ref{PsiDiffeq})
implies the single coproduct relations,
\be
\Psi^w = \Psi^u + \Psi^{1-u} = \Psi^v + \Psi^{1-v} = 0,
\label{Psisinglecop}
\ee
and the additional double coproduct ones,
\bea
\Psi^{1-u,1-v} &=& \Psi^{1-v,1-u} = 0, \label{PsiESlike}\\
\Psi^{u,v} &=& \Psi^{v,u} = - \Psi^{1-u-v,u} = - \Psi^{1-u-v,v} = \Psi^{(L-1)} \,,
\label{Psidoublecop}
\eea
where we have dropped the $(L)$ superscript on $\Psi^{(L)}$ for clarity.
Using these relations, the flip symmetry $\Psi^{(L)}(u,v) = \Psi^{(L)}(v,u)$,
and the vanishing of $\Psi$ in the collinear limit $w = 1-u-v \to0$,
it is easy to locate the pentabox ladder integrals in $\cM$; we have done
so through five loops.

The extended-Steinmann-like constraints we have observed for the form factor
space $\cM$ do not seem to have a direct physical explanation, based on causality,
within that space alone.  On the other hand, when $\Psi^{(L)} \in \cM$
is interpreted as a pentaladder integral in heptagon kinematics, the
constraints follow from the extended Steinmann relations
 in planar $\mathcal{N} = 4$ sYM
theory~\cite{Dixon:2016nkn,Caron-Huot:2019bsq}, which have a causal interpretation.
This connection can be seen most easily by building the full space of
polylogarithms with the alphabet~\eqref{eq:pentaladder_alphabet}---interpreted
as heptagon letters via the
definitions~\eqref{eq:pentaladder_cross_ratios}---that satisfy the first-entry
and extended Steinmann conditions.\footnote{Since the extended Steinmann
  relations are most directly formulated in terms of non-DCI variables,
  their implications in the DCI heptagon alphabet~\eqref{eq:pentaladder_alphabet}
  are not immediately clear; however, they can be imposed easily on an ansatz
  of DCI symbol letters. While a DCI formulation of cluster adjacency does exist,
  it only guarantees that a cluster-adjacent form of the symbol exists in terms
  of $\mathcal{X}$-coordinates, not that every representation of the symbol
  in terms of $\mathcal{X}$-coordinates will obey cluster
  adjacency~\cite{Golden:2018gtk}.}
This construction generates the $\mathcal{P}$ subspace of the form factor space
$\cM$, which has dimension
\be
d_{\mathcal{P},\,\text{symb}}(t)
\equiv \sum_{w=1}^\infty d_{\mathcal{P}} (w) t^w
= \frac{1 - t - 2 t^2 + t^3}{(1-2 t)(1 - t - t^2)}
= 1 +  2 t + 3 t^2 + 6 t^3 + 11 t^4 + \cdots \, 
\ee
at symbol level. Notice that only two functions appear at weight one,
consistent with the above comment that $1-u-v$ does not satisfy the
first-entry condition in heptagon kinematics. Moreover, at weight two,
only the functions
\begin{align}
\Li_2(1-1/u), \quad \Li_2(1-1/v), \quad \ln^2 (u/v) ,
\end{align}
survive after applying the heptagon Steinmann conditions.
As has been previously observed in heptagon kinematics, imposing the
extended Steinmann relations at each higher weight also implies
cluster adjacency~\cite{Drummond:2017ssj,Drummond:2018dfd,Golden:2019kks},
even though the latter condition seems to imply more constraints.

In terms of the labelling of the heptagon cross ratios $u_i$,
and the heptagon $g$ letters described in ref.~\cite{Dixon:2020cnr},
the five letters~(\ref{eq:pentaladder_alphabet}) become
\bea
u &=& u_2 u_6 = g_{1,2} \, g_{1,6} \,, \qquad
v = u_4 u_7 = g_{1,4} \, g_{1,7} \,, \nonumber\\
1-u &=& g_{3,4} \,, \qquad 1-v = g_{3,2} \,, \qquad
1-u-v = g_{4,3} \,.
\label{pentalettersg}
\eea
To restate the previous connection in terms of these letters:
the letters $g_{3,i}$ and $g_{3,i+2}$ are never found adjacent
to each other in the extended Steinmann heptagon function space.

It would be very interesting to find a constructive description of the space
$\mathcal{P}$, perhaps as a prelude to finding one for $\cM$ or $\cC$. 
We leave this to future work.

\section{Near-collinear limit at orders \texorpdfstring{$T^2$}{T**2} and \texorpdfstring{$T^4$}{T**4}}
\label{app:OPEformulas}

In this appendix, we describe further the structure of the near-collinear
expansion of the remainder function discussed in section~\ref{sec:FFOPE},
providing explicit results at two and three loops.

The near-collinear expansion of the three-point remainder function
$R\equiv R_3$ in $T$ has the form,
\be
R^{(L)}(T,S) = \sum_{j=1}^\infty \sum_{k=0}^{L-1}
  T^{2j} \, \ln^k T \, R_{j,k}^{(L)}(S)  \,.
\label{gennearcol}
\ee
From the FFOPE side, the $T^2$ terms are completely determined,
as are essentially all of the $T^4$ terms.  As mentioned
in section~\ref{sec:FFOPE}, the FFOPE results are generated as
a high-order series expansion around $S=0$.  One can construct an
ansatz for the closed-form dependence on $S$ in terms of
HPLs up to a certain weight with suitable rational-function prefactors.
With enough terms in the series expansion, one can fix all of the
coefficients in the ansatz.

Alternatively, it is straightforward to obtain the complete dependence on $S$
by using the coproduct representation to compute the $v$ (or $T$) derivative
of each function in $\cM$ in terms of lower-weight functions.
The integration in $T$ is trivial to do, order-by-order in $T$,
given the expansion~(\ref{gennearcol}).
The HPLs are generated by the $T^0$ term in the expansion,
where the derivative in $u$ has to be integrated up (at least for
the terms with no $\ln T$), but this is also straightforward.
After constructing the near-collinear limits of all the functions,
the results for $E$ or $\EE$, and then for $R$, can be obtained,
in principle to any power of $T^2$.  

Here we provide some terms at order $T^2$ and $T^4$.
It is convenient to take the argument of the harmonic polylogarithms
to be $x = -S^2$.  At order $T^2$, we can separate out the rational
prefactors by writing
\be
R_{1,k}^{(L)} = \frac{(1-x)^2}{x} A^{(L)}_k + (1-x) B^{(L)}_k + C^{(L)}_k \,,
\label{ABCdecomp}
\ee
where $A^{(L)}_k$, $B^{(L)}_k$ and $C^{(L)}_k$ are pure functions, although
not of uniform weight.   This particular decomposition is useful
because the maximum weight of $A^{(L)}_k$ is $2L-k-1$, whereas
$B^{(L)}_k$ and $C^{(L)}_k$ turn out to have a maximal weight of $2L-k-2$.

At two loops, these functions are given by
\bea
A^{(2)}_1 &=& - 8 \, H_{1}(x) \, \ln(-x)
- 16 \, H_{1,1}(x) - 8 \, H_{1}(x) \,,
\label{A21}\\
B^{(2)}_1 &=& 8 \, \ln(-x) \,,
\label{B21}\\
C^{(2)}_1 &=& 8 \,,
\label{C21}\\
A^{(2)}_0 &=&  4 \, \bigl[ 2 \, H_{1,1}(x) + H_{1}(x) \bigr] \, \ln(-x)
+ 8 \, H_{0,0,1}(x) + 16 \, H_{1,1,1}(x) - 8 \, \zeta_2 \, H_{1}(x)
\nonumber\\&&\hskip0.0cm\null
+ 8 \, H_{1,1}(x) + 4 \, H_{1}(x) \,,
\label{A20}\\
B^{(2)}_0 &=& - 4 \, \ln(-x) \,,
\label{B20}\\
C^{(2)}_0 &=& 8 \, \zeta_2 - 12 \,.
\label{C20}
\eea
At three loops, suppressing now the argument $x$ of the HPLs, we have
\bea
A^{(3)}_2 &=& - 12 \, H_{1} \, \ln^2(-x)
- 32 \, \bigl[ 2 \, H_{1,1} + H_{1} \bigr] \, \ln(-x)
+ 8 \, H_{0,0,1} - 128 \, H_{1,1,1} - 8 \, \zeta_2 \, H_{1}
\nonumber\\&&\hskip0.0cm\null
- 64 \, H_{1,1} - 32 \, H_{1} \,,
\label{A32}\\
B^{(3)}_2 &=& 32 \, \ln(-x) \,,
\label{B32}\\
C^{(3)}_2 &=& 12 \, \ln^2(-x) + 8 \, \zeta_2 + 24 \,,
\label{C32}
\eea
\bea
A^{(3)}_1 &=& 12 \, \bigl[ 2 \, H_{1,1} + H_{1} \bigr] \, \ln^2(-x)
+ 4 \, \bigl[ 2 \, H_{1,0,1} + 4 \, H_{0,0,1} + 2 \, H_{0,1,1}
      + 48 \, H_{1,1,1}
\nonumber\\&&\hskip0.5cm\null
      + 6 \, \zeta_2 \, H_{1}
      + 24 \, H_{1,1} + 11 \, H_{1} \bigr] \, \ln(-x)
\nonumber\\&&\hskip0.0cm\null
+ 24 \, H_{1,0,0,1} + 24 \, H_{0,1,0,1} + 48 \, H_{0,0,1,1}
+ 384 \, H_{1,1,1,1}
\nonumber\\&&\hskip0.0cm\null
+ 8 \, \zeta_2 \, \bigl[ 2 \, H_{1,1} + H_{1} \bigr]
- 8 \, \zeta_3 \, H_{1}
+ 24 \, H_{0,0,1} + 192 \, H_{1,1,1}
\nonumber\\&&\hskip0.0cm\null
+ 12 \, H_{0,1} + 96 \, H_{1,1} + 60 \, H_{1} \,,
\label{A31}\\
B^{(3)}_1 &=& 2 \, \ln^2(-x)
+ \bigl[ - 32 \, \zeta_2 - 8 \, H_{1} - 60 \bigr] \, \ln(-x)
+ 24 \, H_{0,1} + 12 \, \zeta_2 \,,
\label{B31}\\
C^{(3)}_1 &=& - 14 \, \ln^2(-x)
+ \bigl[ - 8 \, H_{0,1} + 8 \, \zeta_2 \bigr] \, \ln(-x)
+ 24 \, H_{0,0,1} + 8 \, \zeta_3
- 20 \, \zeta_2 - 96 \,,
\label{C31}
\eea
\bea
A^{(3)}_0 &=& \bigl[ - 24 \, H_{1,1,1} - 2 \, H_{1,0,1}
   + 2 \, H_{0,0,1} - 2 \, H_{0,1,1} - 2 \, \zeta_2 \, H_{1}
   - 12 \, H_{1,1} - 5 \, H_{1} \bigr] \, \ln^2(-x)
\nonumber\\&&\hskip0.0cm\null
+ \bigl[ - 12 \, H_{0,1,1,1} - 8 \, H_{1,0,0,1} - 8 \, H_{0,1,0,1}
    - 16 \, H_{0,0,1,1} - 8 \, H_{1,1,0,1} - 12 \, H_{1,0,1,1}
\nonumber\\&&\hskip0.5cm\null
    - 192 \, H_{1,1,1,1}
    - 12 \, \zeta_2 \, ( 2 \, H_{1,1} + H_{1} ) - 4 \, \zeta_3 \, H_{1}
    - 4 \, H_{1,0,1} - 4 \, H_{0,1,1} - 8 \, H_{0,0,1}
\nonumber\\&&\hskip0.5cm\null
    - 96 \, H_{1,1,1}
    - 4 \, H_{0,1} - 46 \, H_{1,1} - 24 \, H_{1} \bigr] \, \ln(-x)
\nonumber\\&&\hskip0.0cm\null
- 48 \, H_{0,0,1,1,1} - 72 \, H_{0,0,0,0,1} - 12 \, H_{0,0,0,1,1}
- 24 \, H_{1,1,0,0,1} - 24 \, H_{1,0,1,0,1}
\nonumber\\&&\hskip0.0cm\null
- 36 \, H_{1,0,0,1,1}
- 24 \, H_{0,1,1,0,1} - 12 \, H_{1,0,0,0,1} - 12 \, H_{0,1,0,0,1}
- 12 \, H_{0,0,1,0,1}
\nonumber\\&&\hskip0.0cm\null
- 36 \, H_{0,1,0,1,1}
- 384 \, H_{1,1,1,1,1}
+ 16 \, \zeta_2 \, \bigl[ H_{0,0,1} - H_{1,1,1} \bigr]
+ 4 \, \zeta_3 \, \bigl[ 2 \, H_{1,1} + H_{1} \bigr]
\nonumber\\&&\hskip0.0cm\null
+ 77 \, \zeta_4 \, H_{1}
- 12 \, H_{1,0,0,1} - 12 \, H_{0,1,0,1} - 24 \, H_{0,0,1,1}
- 192 \, H_{1,1,1,1}
- 8 \, \zeta_2 \, H_{1,1}
\nonumber\\&&\hskip0.0cm\null
- 6 \, H_{0,0,1} - 6 \, H_{0,1} - 6 \, H_{0,1,1}
- 96 \, H_{1,1,1}
- 54 \, H_{1,1} - 4 \, \zeta_2 \, H_{1}
- 36 \, H_{1} \,,
\label{A30}
\eea
\bea
B^{(3)}_0 &=& 2 \, \ln^2(-x) \, H_{1}
+ \bigl[ - 6 \, H_{0,1} + 4 \, H_{1,1} + 6 \, H_{1}
    + 14 \, \zeta_2 + 36 \bigr] \, \ln(-x)
\nonumber\\&&\hskip0.0cm\null
+ 6 \, H_{0,0,1} - 12 \, H_{0,1} - 12 \, H_{0,1,1}
+ 6 \, \zeta_3 - 6 \, \zeta_2 \,,
\label{B30}\\
C^{(3)}_0 &=& \bigl[ 2 \, H_{0,1} + 2 \, \zeta_2 + 3 \bigr] \, \ln^2(-x)
+ \bigl[ - 8 \, H_{0,0,1} + 4 \, H_{0,1,1} + 2 \, H_{0,1}
    + 4 \, \zeta_3 - 2 \, \zeta_2 \bigr] \, \ln(-x)
\nonumber\\&&\hskip0.0cm\null
- 12 \, H_{0,0,1,1} + 12 \, H_{0,0,0,1} - 6 \, H_{0,0,1}
- 77 \, \zeta_4 - 10 \, \zeta_3 - 6 \, \zeta_2 + 120 \,.
\label{C30}
\eea
The expressions at four and five loops may be found in the ancillary file
{\tt T2terms.txt}.

Similarly, at order $T^4$, the rational prefactors can be separated out
by letting
\bea
R_{2,k}^{(L)} &=& \frac{(1-x)(1+x+3x^2-x^3)}{x^2} D^{(L)}_k
+ \frac{(1-x)^2(1+2x-x^2)}{x^2} E^{(L)}_k
+ \frac{(1-x)^4}{x^2} F^{(L)}_k
\nonumber\\&&\hskip0.0cm\null
+ (1-x) G^{(L)}_k + K^{(L)}_k \,,
\label{DEFGKdecomp}
\eea
where $D^{(L)}_k$, $\ldots$, $K^{(L)}_k$ are pure functions, although again
not of uniform weight.  The maximum weight of $D^{(L)}_k$ is $2L-k-1$,
while the other coefficients have at most one weight lower.

At two loops, these functions are given by
\bea
D^{(2)}_1 &=& - 4 \, H_{1} \, \ln(-x) - 8 \, H_{1,1} + \frac{1}{2} \,,
\label{D21}\\
E^{(2)}_1 &=& \ln(-x) \,,
\label{E21}\\
F^{(2)}_1 &=& - \ln(-x) - 2 \, H_{1} - \frac{1}{2} \,,
\label{F21}\\
G^{(2)}_1 &=& 4 \, \ln(-x) - 2 \,,
\label{G21}\\
K^{(2)}_1 &=& 8 \, \ln(-x) + 1 \,,
\label{K21}
\eea
\bea
D^{(2)}_0 &=& 4 \, H_{1,1} \, \ln(-x) + 8 \, H_{1,1,1} + 4 \, H_{0,0,1}
- \zeta_2 \, \bigl[ 4 \, H_{1} - 3 \bigr] + H_{1} - \frac{1}{8} \,,
\label{D20}\\
E^{(2)}_0 &=& \biggl[ 2 \, H_{1} - \frac{1}{4} \biggr] \, \ln(-x)
- 4 \, H_{0,1} - 2 \, \zeta_2 \,,
\label{E20}\\
F^{(2)}_0 &=& - \biggl[ H_{1} - \frac{1}{4} \biggr] \, \ln(-x)
- 2 \, H_{1,1} - \zeta_2 - \frac{1}{2} \, H_{1} + \frac{1}{8} \,,
\label{F20}\\
G^{(2)}_0 &=& - 5 \, \ln(-x) - 4 \, \zeta_2 - 4 \, H_{1} + \frac{1}{2} \,,
\label{G20}\\
K^{(2)}_0 &=& - 4 \, \ln(-x) - 2 \, \zeta_2 - \frac{7}{4} \,.
\label{K20}
\eea

At three loops, the $\ln^2T$ coefficients are
\bea
D^{(3)}_2 &=& - \biggl[ 6 \, H_{1} - \frac{9}{2} \biggr] \, \ln^2(-x)
- 32 \, H_{1,1} \, \ln(-x) - 64 \, H_{1,1,1} + 4 \, H_{0,0,1}
\nonumber\\&&\hskip0.0cm\null
- \zeta_2 \, ( 4 \, H_{1} - 3 ) + \frac{3}{2} \,,
\label{D32}\\
E^{(3)}_2 &=& - 3 \, \ln^2(-x) - ( 4 \, H_{1} - 3 ) \, \ln(-x)
- 4 \, H_{0,1} - 2 \, \zeta_2 \,,
\label{E32}\\
F^{(3)}_2 &=& - \frac{3}{2} \, \ln^2(-x) - ( 8 \, H_{1} + 3 ) \, \ln(-x)
- 16 \, H_{1,1} - 6 \, H_{1} - \zeta_2 - \frac{3}{2} \,,
\label{F32}\\
G^{(3)}_2 &=& - 6 \, \ln^2(-x) + 12 \, \ln(-x) - 4 \, \zeta_2 - 6 \,,
\label{G32}\\
K^{(3)}_2 &=& - 3 \, \ln^2(-x) + 8 \, \ln(-x) - 2 \, \zeta_2
- \frac{1}{2} \,.
\label{K32}
\eea
The remaining three-loop $T^4$ coefficients, and the four- and five-loop ones,
can be found in the ancillary file {\tt T4terms.txt}.


\bibliographystyle{JHEP}
\bibliography{ff}

\end{fmffile}
\end{document}